\documentclass[reprint,showkeys,aps,prx,floats,twocolumn,
superscriptaddress,noeprint,fontsize=10,longbibliography,nofootinbib]{revtex4-1}
\usepackage[english]{babel}
\usepackage[utf8]{inputenc}
\usepackage{graphicx,epsfig}
\usepackage{graphics,dcolumn,bm,epic, eepic}
\usepackage{amssymb,amsmath,amsfonts,multirow,rotate,color}
\usepackage{soul}
\usepackage{footnote}
\usepackage{setspace}

\usepackage{makecell}

\usepackage[colorinlistoftodos, color=green!40, prependcaption]{todonotes}
\usepackage{amsthm}
\usepackage{mathtools}
\usepackage{physics}
\usepackage{xcolor}
\usepackage{hyperref}
\usepackage{placeins}
\hypersetup{	
    colorlinks=true,
    linkcolor={blue!60!black},
    citecolor={blue!80!black},
    urlcolor={blue!80!black}
}
\usepackage{adjustbox}
\usepackage{placeins}
\usepackage{lipsum}
\usepackage{csquotes}

\AtBeginDocument{%
    \newwrite\bibnotes
    \def\bibnotesext{Notes.bib}
    \immediate\openout\bibnotes=\jobname\bibnotesext
    \immediate\write\bibnotes{@CONTROL{REVTEX41Control}}
    \immediate\write\bibnotes{@CONTROL{%
            apsrev41Control,author="08",editor="1",pages="1",title="0",year="1"}}
    \if@filesw
    \immediate\write\@auxout{\string\citation{apsrev41Control}}%
    \fi
}%

\definecolor{darkred}{rgb}{0.8, 0.0, 0.0}

\newcommand{\newtext}[1]{\textcolor{black}{#1}}

\newcommand{\nocontentsline}[3]{}
\newcommand{\tocless}[2]{\bgroup\let\addcontentsline=\nocontentsline#1{#2}\egroup}

\begin{document}
    \title{Unveiling the higher-order organization of multivariate time series}
    \author{Andrea Santoro}
    \affiliation{Neuro-X Institute, \'{E}cole Polytechnique F\'{e}d\'{e}rale de Lausanne, 1202 Geneva, Switzerland}
    \author{Federico Battiston}
    \affiliation{Department of Network and Data Science, Central European University, 1100 Vienna, Austria}
    \author{Giovanni Petri}
    \affiliation{CENTAI, Corso Re Inghilterra 3, Turin, Italy}
    \author{Enrico Amico}
    \thanks{Corresponding author: \href{mailto:enrico.amico@epfl.ch}{enrico.amico@epfl.ch}}
    \affiliation{Neuro-X Institute, \'{E}cole Polytechnique F\'{e}d\'{e}rale de Lausanne, 1202 Geneva, Switzerland}
    \affiliation{Department of Radiology and Medical Informatics, University of Geneva, 1211 Geneva, Switzerland}

    \date{\today}
    
    \begin{abstract}
        Time series analysis has proven to be a powerful method to characterize several phenomena in biology, neuroscience and economics, and to understand some of their underlying dynamical features. Despite a plethora of methods have been proposed for the analysis of multivariate time series, most of
        them \newtext{neglect the effect of non-pairwise interactions on the emerging dynamics.}
        Here, we propose a novel framework to characterize the temporal evolution of higher-order dependencies within multivariate time series. 
        Using network analysis and topology, we show that, \newtext{unlike traditional tools based on pairwise statistics}, our framework robustly differentiates various spatiotemporal regimes of coupled chaotic maps, including chaotic dynamical phases and various types of synchronization. 
        \newtext{Hence, using the higher-order co-fluctuation patterns in simulated dynamical processes as a guide, we highlight and quantify signatures of higher-order patterns in data from brain functional activity, financial markets, and epidemics.}  
        Overall, our approach sheds new light on the higher-order organization of multivariate time series, allowing a better characterization of dynamical group dependencies inherent to real-world data.
    \end{abstract}
    
    \keywords{multivariate time series, higher-order dependencies, topological data analysis, complex systems, network theory}
    
    \maketitle
    
    \noindent
    The growing availability of rich, often temporally resolved, data coming from many different complex systems, has led to the possibility of studying in detail their behaviour and --often-- their internal mechanisms.
    Examples of such systems include epidemics, social contagion, as well as financial, brain, and biological signals. 
    All of these systems are composed by large numbers of elementary units interacting in heterogeneous fashion with each other, and --in virtually all cases-- displaying emergent properties at the macroscopic level. 
    Due to the crucial importance of the patterns of interactions composing such systems, it is no surprise that complex networks emerged as a powerful framework to investigate the structure and dynamics of such systems~\cite{albert2002statistical,boccaletti2006complex}, and have helped to characterize several real-world phenomena, including disease spreading~\cite{pastor-satorras2015epidemic}, synchronisation~\cite{arenas2008synchronization}, diffusion~\cite{barrat2008dynamical}, and opinion formation~\cite{watts2007influentials}. 
    
    Despite being widely considered as the reference model for many real-world complex systems~\cite{barabasi2016network,latora2017book,newman2018networks}, networks are limited to describing interactions between two units (or nodes) at a time. 
    This however clashes with the growing empirical evidence for group interactions in social systems~\cite{benson2016higherorder}, neuroscience~\cite{petri2014homological,giusti2015clique,sizemore2018cliques}, ecology~\cite{grilli2017higherorder} and biology~\cite{sanchez-gorostiaga2019highorder}. In all the aforementioned cases, connections and relationships do not take place only between pairs of nodes, but also as collective actions of groups of nodes. 
    By taking into account the higher-order (group) interactions in more refined models, such as hypergraphs and simplicial complexes~\cite{battiston2020networks,torres2021why}, several recent studies have shown that
    the presence of higher-order interactions can have a substantial impact
    on the dynamics of interacting systems~\cite{battiston2021physics,battiston2022higher}, ranging from alterations of the synchronization~\cite{millan2019synchronization} and diffusion~\cite{schaub2020random, carletti2020random} properties, to new collective dynamics in  social~\cite{iacopini2019simplicial,dearruda2020social,sahasrabuddhe2021modelling} and evolutionary processes~\cite{alvarez-rodriguez2021evolutionary}. 
    
    Yet, direct measurements of pairwise or group interactions to inform and constrain such higher-order models are rarely available. Hence, one must typically rely on indirect data, commonly extracted from time series of node activities, under the assumption that the system's repertoire of spatiotemporal activity patterns encodes information about the underlying interactions. 
    Indeed, examples of these complex patterns are observed in the neuronal activity of the brain, supporting a wide variety of motor and cognitive functions~\cite{deco2011emerging,avena-koenigsberger2018communication}, in financial markets, where partially synchronized patterns often reflect periods of financial stress~\cite{mantegna1999introduction,peron2011collective}, but also in the co-evolution of biological species~\cite{olesen2008temporal,friedman2017community,ebert2020host}.  
    
    While the inference of pairwise interactions has a long history~\cite{brugere2018network}, researchers have only recently taken the first steps towards reconstructing or filtering higher-order interactions~\cite{young2021hypergraph,musciotto2021detecting,wang2022full,lizotte2022hypergraph}. \newtext{In particular, methods relying only on pairwise statistics might be in principle insufficient as significant information can be present only in the joint probability distribution and not in the pairwise marginals, therefore failing at identifying higher-order behaviours~\cite{rosas2022disentangling}.} To date, it remains unclear to what degree the information encoded in multivariate time series stems from independent individual entities or, rather, from their group interactions. A clear example of this issue is provided by the conventional ``functional connectivity'' between two brain regions~\cite{bullmore2009complex,sporns2010networks}: a pairwise connection is drawn irrespective of whether the activities of the two regions peaked as a pair, or as part of a larger group of functionally coherent regions.
    
    Existing proposals to address this issue are \newtext{mostly} limited in their capacity to describe either the temporality or the complexity of such higher-order interactions, \newtext{with only few exceptions on the topic~\cite{faes2022framework}}. For example, a recent set of information-theoretic methods characterized higher-order dependencies in multivariate time series by quantifying the intrinsic statistical synergy and redundancy in groups of three or more interacting variables~\cite{rosas2019quantifying,rosas2020reconciling,gatica2021highorder,stramaglia2021quantifying}. Moreover, a recent approach~\cite{macmahon2015community} at the interface of network science and random matrix theory has also proven suitable to unveil the mesoscopic organization of correlation matrices. Yet, while very powerful, these methods hardly capture the information about the dynamics of the system, because they require integration over time. \newtext{A recent exception comes from a work introducing a spectral decomposition that resolve the statistical synergy and redundancy of groups of variables into different frequency bands, which allows to analyze locally in time higher-order dependencies~\cite{faes2022framework}}. By contrast, tools coming from network neuroscience and signal processing easily deal with the dynamics of multivariate time series~\cite{tagliazucchi2012criticality,liu2013timevarying,karahanoglu2015transient,preti2017dynamic,liu2018coactivation,faskowitz2020edgecentric, esfahlani2020highamplitude,vandeville2021when}, but only focus on pairwise statistics and neglect the effects of higher-order interactions. As a result, a principled approach to quantify the instantaneous dynamics of groups of nodes \newtext{and possibly infer its higher-order representation} is still missing.
    
    \begin{figure*}[t!]
        \centering
        \includegraphics[width=0.95\textwidth]{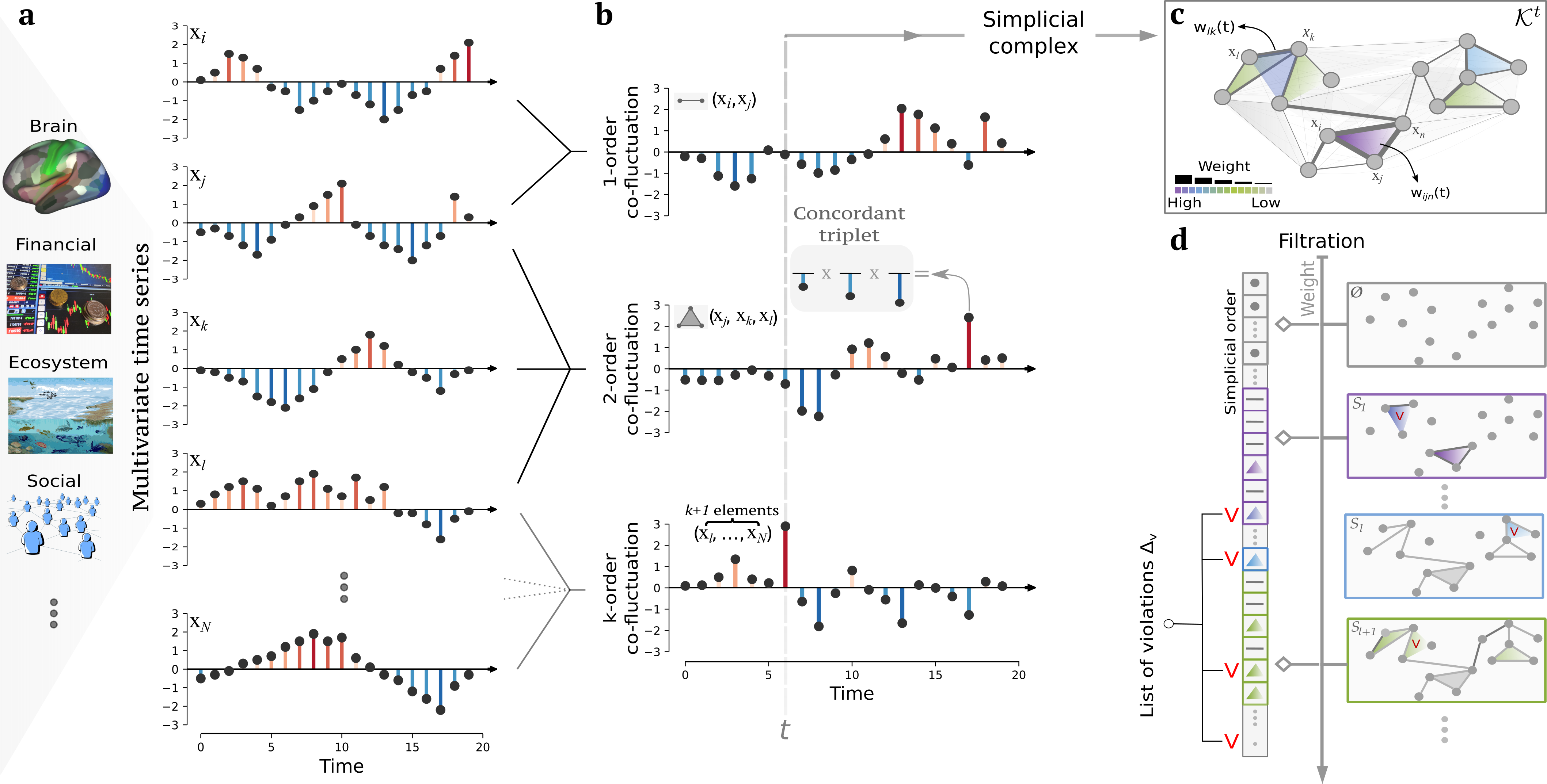}
        \caption{\textbf{Unveiling the higher-order structure of multivariate time series: Schematic representation.} \textbf{(a)} We start by extracting different co-fluctuation patterns according to their order from the ``raw'' nodal time series. \textbf{(b)} The generic element of a $k$-order co-fluctuation pattern is calculated by $z$-scoring each time series, and performing an element-wise product of $k+1$ time series. We further $z$-score each of these group product time series so that the magnitude of the time-resolved co-fluctuation is comparable across group size (i.e. pairs, triangles, etc.). To distinguish concordant group interactions from discordant ones, we impose that concordant signs are always positively mapped, while discordant signs are negatively mapped.  \textbf{(c)} For each time frame $t$, a weighted simplicial complex is constructed by merging together all the k-order co-fluctuations. \textbf{(d)} Finally, a weight filtration made of all the k-order co-fluctuations allows to identify weighted holes when k-order patterns are gradually included. For simplicity, we set $k=2$ for panels \textbf{(c, d)}.
        }
        \label{fig:fig1_framework}
    \end{figure*}

    Here, we \newtext{propose} a novel framework to characterize the instantaneous co-fluctuation patterns of signals at all orders of interactions (pairs, triangles, etc.), and to investigate the global topology of such co-fluctuations. We do this by bridging time series analysis, complex network theory, and topological data analysis~\cite{wasserman2018topological}.
    We first validate the framework by exploring the rich high-dimensional dynamics displayed by canonical models of spatiotemporal chaos. In particular,  we demonstrate that, unlike traditional tools of time-series analysis \newtext{based on pairwise statistics}~\cite{wei2005time,wei2019multivariate,zou2019complex}, higher-order measures are able to reveal the subtleties of different spatiotemporal regimes at the level of individual frames. We then use the insights obtained from these synthetic models as a Rosetta stone to interpret the higher-order structures reconstructed from time series concerning three diverse real-world case studies: resting-state brain activity (as measured by fMRI data), stock option prices, and epidemiological incidence of various diseases in the United States. 
    In all cases, we unveil \newtext{additional} rich higher-order information that is not captured at the node and dyadic network level, and highlight distinct topological dynamical regimes, which in turn yield to instantaneous classification of the system's states.\\
    Finally, we show how the inferred \newtext{dynamical} higher-order structure \newtext{provides instantaneous topological snapshots of the spatial configuration of the system}, which can be used as input for further tasks on real datasets, \newtext{including} detecting local integration of brain regions, exploring periods of financial crisis, or classifying disease type from spatial spreading patterns. 
    
    \tocless{\section*{Results}}
    \noindent
    \textbf{Topological markers of higher-order structure in multivariate time series.} 
    Simplicial complexes are well suited as modelling framework to describe the co-existence of pairwise and higher-order interactions~\cite{battiston2020networks}.  
    In its most basic definition, a $k$-simplex $\sigma$ is a set of $k+1$ vertices $\sigma = [p_0, \ldots, p_k ]$. 
    A collection of simplices is a simplicial complex $\mathcal{K}$ if for each simplex $\sigma$ all its possible subfaces (defined as subsets of $\sigma$) are themselves contained in $\sigma$ (see Methods and Ref.~\cite{hatcher2005algebraic} for details). 
    Via this representation it is then easy to distinguish between a group interaction among three elements, which can be represented as a 2-simplex (or ``filled'' triangle) $[p_0, p_1, p_2]$, and the three pairwise interactions between the nodes, that is, the collection of 1-simplices (edges) $[p_0, p_1], [p_0, p_2], [p_1, p_2]$.  
    The relative importance of pairwise versus higher-order interactions can be encoded in weights over the simplices, resulting in the so-called \textit{weighted simplicial complexes}. 
    
    We rely on this representation to describe the higher-order dependencies among multiple time series. 
    Our approach can be summarised in five main steps: 
    First, \textit{i)} we $z$-score the $N$ original time series (Fig.~\ref{fig:fig1_framework}a), and then \textit{ii)} we calculate the element-wise product of the $z$-scored time series for all the $\binom{N}{k}$ $k$-order patterns (i.e. edges, triangles, etc.). 
    Here, the generic elements represent the instantaneous co-fluctuation magnitude between a $(k+1)$ group interaction. Then, \textit{iii)} the resulting new set of time series encoding the $k$-order co-fluctuations (which corresponds to the so-called \textit{edge time series}~\cite{faskowitz2020edgecentric} in the case of $1$-order co-fluctuation) are then further $z$-scored across time, to make products comparable across $k$-orders (Fig.~\ref{fig:fig1_framework}b). 
    At this point a choice on how to assign signs to the resulting weights is required in order to distinguish fully concordant group interactions (all positive or negative fluctuations) from discordant ones (a mixture of positive and negative fluctuations) in a $k$-order product. Indeed, these two scenarios might end up having similar co-fluctuation z-scored values after a $k$-order product (with $k\geq 2 $), even if they clearly represent different regimes of group synchronization. Hence, we opted to assign positive signs to the fully concordant group interactions, and negative signs to the discordant ones.
    The rationale behind the concordant mapping is such that any simultaneous increased (or decreased) activity relative to baseline $-$ no matter the order of the co-fluctuation $-$ is always marked as positive, therefore reflecting a synchronous co-activation pattern. This adjustment is particularly important for the simplicial filtration step, as detailed below.
    Next, \textit{iv)} for each time frame $t$, we condense all the instantaneous $k$-order co-fluctuations in a single mathematical object, i.e. a weighted  simplicial complex $\mathcal{K}^t$(Fig.~\ref{fig:fig1_framework}c). 
    Lastly, for each time $t$, \textit{v)} we construct a filtration $\mathbb{F}(\mathcal{K}^t)$~\cite{petri2013topological}, i.e. a sequence of simplicial complexes $\emptyset=\mathcal{S}_{0} \subset \mathcal{S}_1 \subset \ldots \subset \mathcal{S}_l \subset \ldots \subset \mathcal{S}_n \subset \mathcal{K}^t$ by sorting all the $k$-order co-fluctuations by their weights (see Methods for details). 
    The filtration proceeds in a top down fashion from larger weights to smaller weights $-$ in the spirit of persistent homology~\cite{edelsbrunner2000topological, zomorodian2005computing, petri2013topological} $-$ so that when $k$-order simplices are gradually included, topological holes start to appear in the simplicial complexes of $\mathbb{F}$ and then potentially close (i.e., descending from more coherent patterns to less coherent). 
    Yet, to maintain valid simplicial complexes as each step of the filtration, only $k$-order simplices respecting the simplicial closure condition can be included. That is, simplices whose subfaces are already contained in the simplicial complex at the previous step. 
    To preserve this property, whenever we would add a simplex that does not satisfy this requirement (e.g. a triangle entering the complex before its edges), we consider it as a \textit{simplicial violation}, and exclude it from the filtration. Note that such violating simplices can be considered as \textit{hyper coherent} structures, as their co-fluctations are stronger than those of its subcomponents~(see Fig.~\ref{fig:fig1_framework}d and Methods for details).
    Note further that in this paper we present results when considering $k=2$, so we take into account simplices only up to triangles. 
    Nevertheless, our framework generalizes naturally to higher orders (i.e. $k\geq 3$). 
    
    In summary, for each time $t$, our framework produces two different outputs:\\
    \begin{enumerate}
        \item  a list of violating triangles,  $\Delta_v=\{(i,j,k),w_{ijk}\}$, induced by the simplicial closure condition; these are 2-simplices (triangles) whose weights co-fluctuate more than at least one of their corresponding 1-simplices (edges).  
        Intuitively, these triangles reflect higher-order states that cannot be merely captured by pairwise co-fluctuations. 
        We then define the \textit{hyper coherence indicator}, as the fraction of violating coherent triangles (i.e. violating triangles with a weight greater than zero) over all the possible coherent triangles (i.e. triangles with a weight greater than zero). 
        \item the simplicial filtration $\mathbb{F}$, a sequence of embedded simplicial complexes $-$ sorted according to coherent patterns $-$ starting with the empty complex and ending with the entire simplicial complex (see right panel of Fig.~\ref{fig:fig1_framework}d). We then compute persistent homology of $\mathbb{F}$ to characterize the persistency of certain topological features (connected components, 1-dimensional cycles, 3D-cavities, etc.)~\cite{zomorodian2005computing,ghrist2008barcodes}.  Here, we focus on examining the 1D cycles in the filtration, i.e. the persistent generators of the first homology group $H_1$, which provide insights about where and when higher synchronised regions emerge. 
        The classical output of persistent homology is a barcode (or equivalently, a persistence diagram), which is a compressed summary describing how long 1D cycles live along $\mathbb{F}$ (see SI Fig.~S1). 
        We rely on this object and define the \textit{hyper complexity indicator} as the Wasserstein distance~\cite{carriere2017sliced} between the persistence diagram of $H_1$ and the empty persistence diagram, corresponding to a space with trivial $H_1$ homology (see Methods for details). 
        We obtain in this way a measure of the topological complexity of the landscape of coherent and incoherent co-fluctuations.
        
    \end{enumerate}

    \begin{figure*}[th!]
        \centering
        \includegraphics[width=0.985\textwidth]{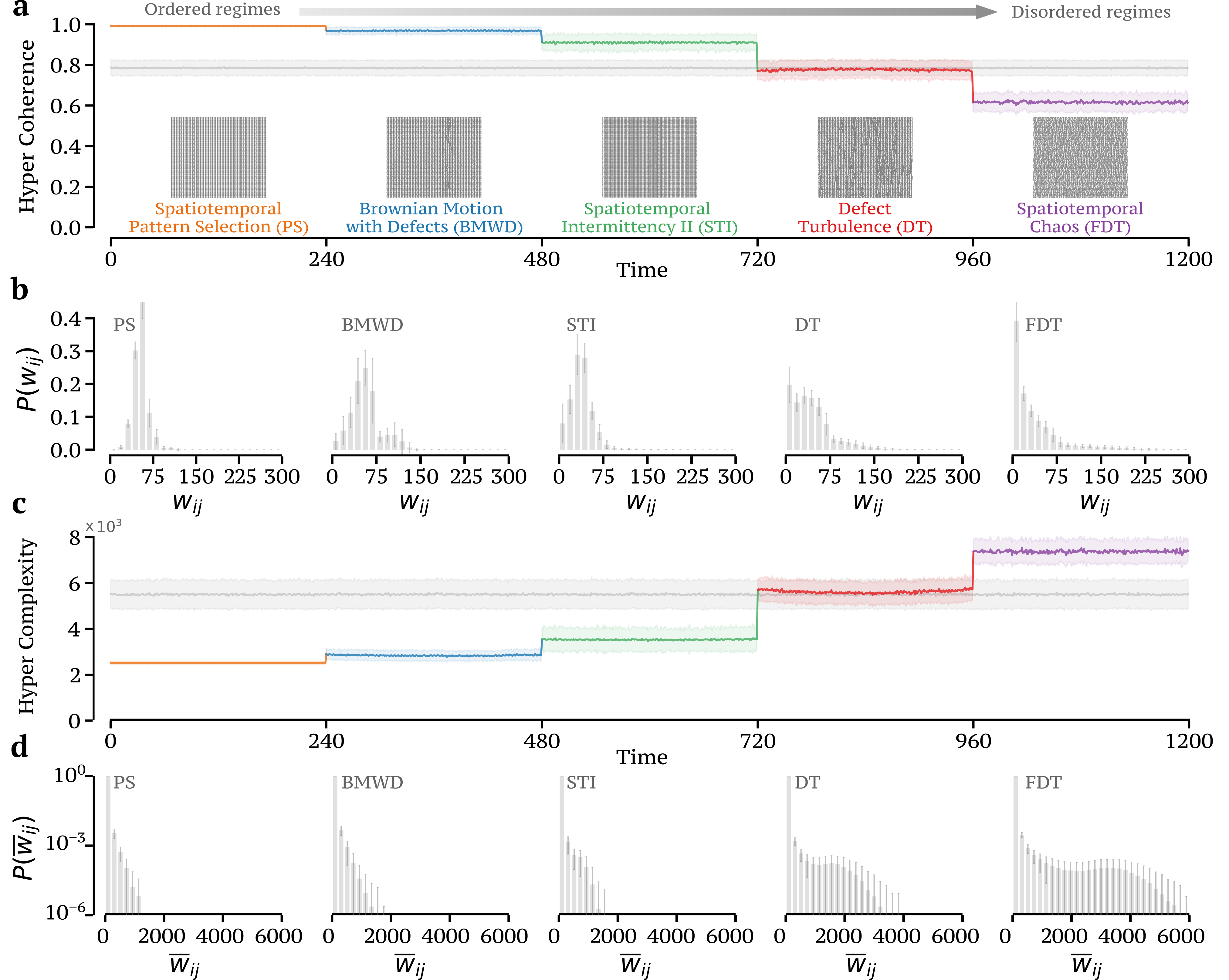}
        \caption{\textbf{Global and local higher-order indicators distinguish the dynamical regimes of coupled chaotic maps.} (\textbf{a}) We report the temporal evolution of the hyper coherence indicator for a multivariate time series with $N=119$ nodes and $T=1200$, obtained by concatenating five different CML regimes with fixed time length  $L=240$. \newtext{Namely, from order to disorder, Pattern Selection (PS) at $\varepsilon= 0.12$, Brownian motion with Defects (BMWD) at $\varepsilon=0.08$, Spatiotemporal Intermittency II (STI) at $\varepsilon = 0.3$, Defect Turbulence (DT) at $\varepsilon = 0.068$, and Fully Developed Turbulence (FDT) at $\varepsilon = 0.05$~\cite{kaneko1989pattern}.} (\textbf{b}) Notably, when projecting the list of violating triangles $\Delta_v$ as a weighted graph (see Methods for the definition of downward projections), the edge weight distribution $P(w_{ij})$ reflects the nature \newtext{and the ``ranking''} of the different dynamical regimes from ordered to disordered. 
            (\textbf{c}) We plot the temporal evolution of the hyper complexity indicator and the (\textbf{d}) distribution of weights $P(\bar{w}_{ij}$) of the homological scaffold constructed from the persistent homology generators of $H_1$~\cite{petri2014homological}. For comparison, we also report in panels (\textbf{a},\textbf{c}) the same indicators for a null model obtained when \newtext{reshuffling without any constraint} the multivariate time series (grey curve). \newtext{See SI Section S4 for the behaviour of the higher-order indicators in more conservative null models.}  Shaded regions and error bars represent standard deviations across 100 independent realizations. }
        \label{fig:fig2_synthetic_results}
    \end{figure*}

    \noindent
    \textbf{Global and local topological markers classify different dynamical regimes.} To gain insights into the performance of our topological indicators, we show here that hyper coherence and hyper complexity easily distinguish different dynamical regimes generated by canonical models of spatiotemporal chaos. 
    As a case study, we consider diffusively coupled map lattices (CMLs)~\cite{kaneko1992overview}, which are high-dimensional dynamical systems defined on discrete time and space, with continuous state variables.
    CMLs are broadly used to model complex spatiotemporal dynamics in several different fields including biology~\cite{bevers1999numerically}, and finance~\cite{hilgers1997turbulent,huang2015heterogeneous}. 
    In particular, we consider a ring lattice with $N$ sites, and we assume that the dynamical evolution of the system of the state $x_i$ of each site $i$ is the result of two different competing dynamics: an internal chaotic dynamic, and an external diffusive coupling dynamic among the first nearest-neighbour sites. 
    Their dynamics can be expressed as
    \begin{equation*}
        x_i(t+1)= (1-\varepsilon) f[x_i(t)]+ \frac{\varepsilon}{2}\left(f\left[x_{\newtext{i-1}}\left(t\right)\right]+f[x_{\newtext{i+1}}(t)]\right)
        \label{eq:kaneko_maps}
    \end{equation*}
    where $i=1,2, \ldots, N$, $\varepsilon \in [0,1]$ is the coupling parameter (or coupling strength), and $f(x)$ is generally a chaotic map. 
    In all our simulations, we have considered the logistic map, i.e. $f(x) = 1 - 1.75\, x^2$.  
    It is now well established~\cite{kaneko1989pattern,kaneko2011complex} that by changing the values of the coupling strength $\varepsilon$ of the chaotic map, CMLs exhibit a great variety of spatiotemporal patterns, including different degrees of synchronization and dynamical phases such as Fully Developed Turbulence (FDT, a phase with incoherent spatiotemporal chaos and high dimensional attractors), Pattern Selection (PS, a phase with suppression of chaos in favour of randomly selected periodic attractor, reflecting quasiperiodic behaviours), and different forms of spatiotemporal Intermittency (STI, chaotic pseudo-phases with low dimensional attractors that interpolate between FDT
    and PS). 
    Together with the latter STI class, we also highlight two other different phases such as Brownian Motion with Defects (BMWD, a phase where defects exist in the system and fluctuate chaotically akin to Brownian motion), and \newtext{Defect Turbulence} (DT, a phase where many defects are generated and turbulently collide together)~\cite{lacasa2015network}. 
    It is worth remarking that the origin of this very rich phase diagram comes from the interplay between the local tendency towards inhomogeneity, which is induced by the chaotic dynamic of each single state, and the global tendency to homogenise the system in space, which is induced by the diffusion dynamic~\cite{lacasa2015network}.
    
    In Figure~\ref{fig:fig2_synthetic_results} we summarize the results of our higher-order approach when applied to these synthetic multivariate series with $N=119$ nodes and $T=1200$, obtained by concatenating five different dynamical phases of CMLs with fixed time length  $L=240$.  
    \newtext{Namely, from order to disorder, PS at $\varepsilon= 0.12$, BMWD at $\varepsilon=0.08$, STI at $\varepsilon = 0.3$, DT at $\varepsilon = 0.068$, and FDT at $\varepsilon = 0.05$ for which a transient of $10^5$ time points has been removed.} A sample of such multivariate time-series is reported in \newtext{SI Fig.~S2, while we study the effect of the $z$-score in SI Section S2.2 and SI Fig.~S3.}  Remarkably, the global hyper coherence indicator reported in Fig.~\ref{fig:fig2_synthetic_results}a clearly distinguishes the different dynamical phases of the CMLs, \newtext{while also preserving the ranking between ordered and disordered states.}
    More precisely, it assigns high values to fully and partially synchronized regimes while, on the contrary, chaotic or turbulent regimes exhibit lower values of hyper coherence. 
    While this indicator provides only global information, refined information can be obtained by projecting the magnitudes of the list of violating triangles $\Delta_v$ as a weighted graph (see Methods for the definition of downward projections).  Also in this case, in fact, the edge weight distribution $P(w_{ij})$ reflects the nature \newtext{and the ``rank''} of the different dynamical regimes (Fig.~\ref{fig:fig2_synthetic_results}b). Periodic series, such as PS, convert into well-peaked distributions, akin to Poisson distributions. By contrast, as disorder enters in the pseudo-phases of the multivariate time series, the edge weight distribution gradually changes its shape, with the limit case of the FDT chaotic series converging towards a fat-tailed distribution.
    
    Similar conclusions can be reached when investigating the temporal evolution of the hyper complexity (Fig.~\ref{fig:fig2_synthetic_results}c). 
    We found, however, some notable differences. 
    For complexity, the lowest value is assigned to periodic patterns (e.g. PS), as these regimes require a low amount of information to be described. 
    Contrarily, chaotic states such as FDT display the highest hyper complexity values. While also this higher-order indicator is able to differentiate the different dynamical regimes of CMLs,  one might assume that the hyper coherence and hyper complexity indicators provide equivalent information as indicated by the strong negative correlation (i.e. Spearman's rank correlation  $\rho \approx -0.95$). 
    We will show in the next section that this is not true in general for real-world multivariate time series. 
    
    Finally, in Fig.~\ref{fig:fig2_synthetic_results}d  we report the edge weight distributions of the persistence homological scaffolds, graphs constructed from the persistent homology generators of $H_1$~(see Methods and Ref.~\cite{petri2014homological} for details). These distributions quantify the topological importance of edges in the co-fluctuation landscape in terms of how persistent are the homological generators to which they belong. Notably, also these distributions change their overall shape as we move from periodic to chaotic multivariate time series \newtext{preserving, also in this case, the rank between order and disorder.}
    
\begin{figure}[t!]
        \centering
        \includegraphics[width=0.475\textwidth]{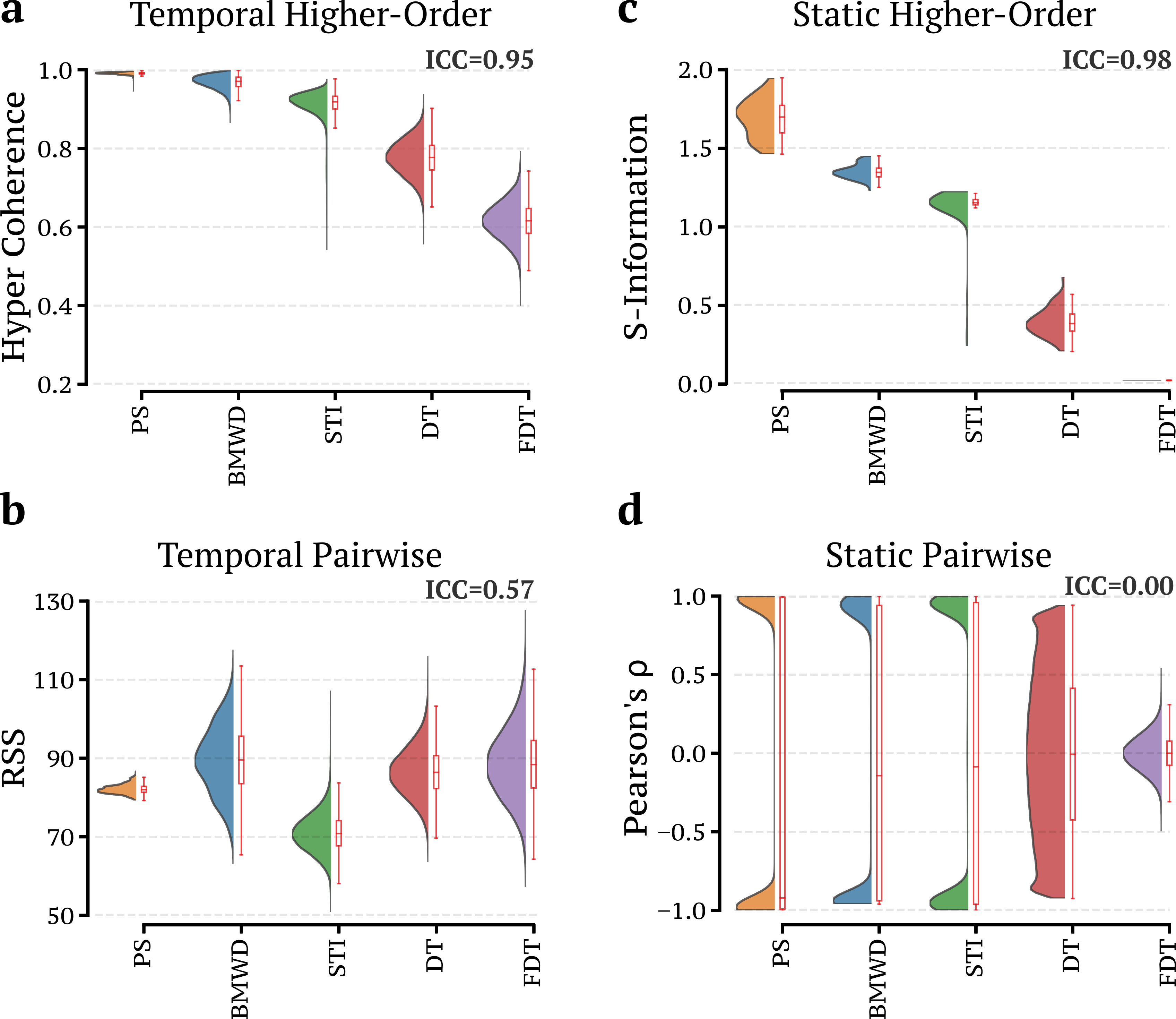}
        \caption{\newtext{\textbf{Higher-order approaches perform better in distinguishing the CML regimes}. For the five dynamical regimes, we report the violin plots of the (\textbf{a}) Hyper Coherence, (\textbf{b}) RSS~\cite{esfahlani2020highamplitude}, (\textbf{c}) S-information~\cite{rosas2019quantifying,gatica2021highorder}, and (\textbf{d}) Person's correlation distributions. Notably, only higher-order approaches are able to distinguish the five dynamical regimes (i.e. $ICC > 0.9$). Also notice that static approaches can only be used when having prior knowledge on the position of each block.}}
        \label{fig:fig3_new}
    \end{figure}

    These results qualitatively confirm that both the global and local topological information extracted with our approach well \newtext{discriminate} among the different dynamical regimes. 
    \newtext{We quantitatively assess the capacity of our higher-order indicators to differentiate between dynamical regimes with the intraclass correlation coefficient (ICC)~\cite{mcgraw1996forming,shrout1979intraclass}, a statistical measure commonly used to determine the agreement between units (or ratings/scores) of different groups.  
        In other words, the ICC describes how strongly units in the same group resemble each other, so that the stronger the agreement, the higher its ICC value. 
        In Figure~\ref{fig:fig3_new}, we report the comparison of several approaches when trying to differentiate the five dynamical regimes of CML.
        From a theoretical standpoint, it is interesting to notice from Figure~\ref{fig:fig3_new}a that the five CMLs regimes map into distinct hyper coherence distributions mirroring some peculiar features intrinsic to each individual spatiotemporal regime.  For instance, apart from a well-defined bulk, the STI regime exhibits a long tail towards lower hyper coherence values, which captures the rare appearance of short chaotic bursts arising from the mismatching of the dynamical phases~\cite{kaneko1989pattern,nunez2013horizontal}.\\
        We find that both our higher-order measures (in Figure \ref{fig:fig3_new} only hyper coherence is shown) have high ICC values (i.e. approximatively $0.95$ and $0.96$ for hyper coherence and hyper complexity, respectively). 
        For a commonly used temporal low-order measure, the RSS statistic~\cite{esfahlani2020highamplitude}, which accounts for the magnitude of peak amplitude in all the 1-order co-fluctuations, i.e. the edge time series (see Methods for a formal definition), we find instead considerably smaller ICC values ($ICC \approx 0.57$) as shown in Figure~\ref{fig:fig3_new}b, implying that the higher-order effects are dominant.  
        Indeed, our dynamic approach is comparable to ``static'' higher-order information-theoretic approaches~\cite{rosas2019quantifying,gatica2021highorder}, i.e. for $s$-information (shown in Figure~\ref{fig:fig3_new}c) we find $ICC \approx 0.98$. However, these quantities are typically computed on temporal windows, while in the topological framework presented here it is possible to have instantaneous information. 
        Finally, as expected, higher-order measures (both static and dynamic) consistently outperform static lower-order methods based on Pearson's correlation, i.e. $ICC \approx 0$ and shown in Figure~\ref{fig:fig3_new}d.
        As a matter of fact, it appears clear that only higher-order approaches are able to effectively distinguish the various spatiotemporal regimes, while lower-order statistics are not able to capture the subtle differences between dynamical states. 
        For detailed comparisons with other ``static'' higher-order and pairwise approaches~\cite{sporns2010networks,macmahon2015community,esfahlani2020highamplitude}, see also SI Section S3 and SI Figures~S6-S7.}
    
    \smallskip
    \noindent
    \textbf{Real-world complex systems exhibit non-trivial hyper coherence structure.}
    \newtext{As examples of applications to the analysis of real-world multivariate time series}, we report the results of the higher-order framework on fMRI signals from the Human Connectome Project (HCP)~\cite{vanessen2013wuminn}, on prices of financial assets in the \newtext{New York Stock Exchange (NYSE)}, and on historical data of several infectious diseases in the US~\cite{scarpino2019predictability,vanpanhuis2013contagious}.
    
    For human brain data, we consider resting-state fMRI signals of the HCP 100 unrelated subjects, employing a cortical parcellation of 100 brain regions~\cite{schaefer2018localglobal} and 19 sub-cortical ones as provided by the HCP release~\cite{glasser2013minimal}, for a total of $N=119$ \newtext{Regions of Interest (ROIs)}. For financial time series, \newtext{we analyse the daily time evolution of $N = 119$ stock prices of some of US companies from the NYSE over the period 2000–2021}. Finally, for the epidemic dataset, we investigate the weekly number of cases at the US state-level ($N=50$) for chlamydia, gonorrhea, influenza, measles, mumps, polio, and pertussis \newtext{(see Methods for details on the datasets)}.

    \begin{figure*}[t!]
        \begin{center}
            \includegraphics[width=0.98\textwidth]{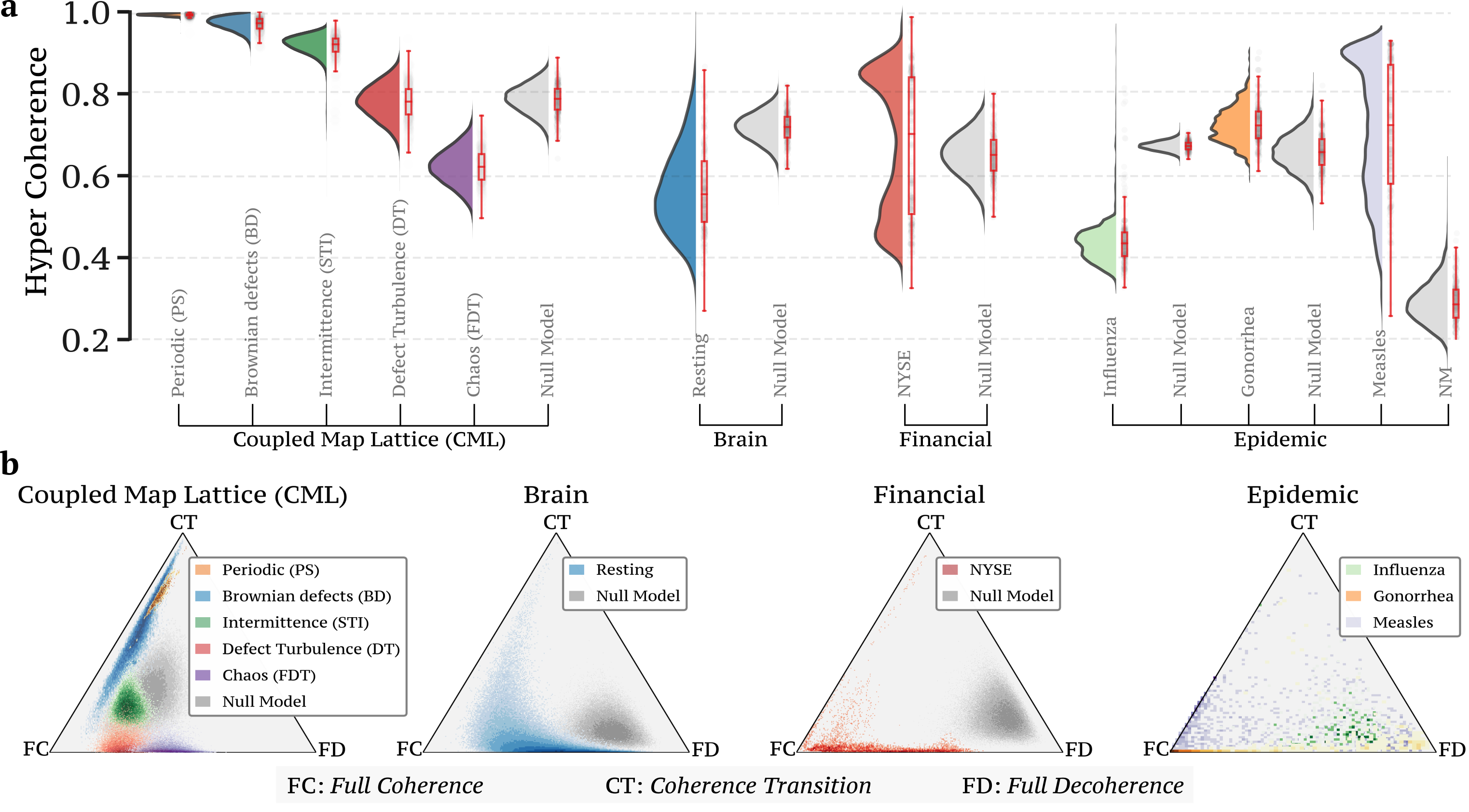}
            \caption{\textbf{Higher-order indicators for real-world multivariate time series}. (\textbf{a}) Violin plots showing the distribution of hyper coherence for three real-world datasets, namely, resting-state fMRI data (N=119 brain regions), financial prices of 119 assets in NYSE, and the US historical data of several infectious diseases at the US state-level (N=50). The real distributions are compared against the five CML dynamical regimes, as well as the corresponding null models obtained when \newtext{independently reshuffling without any constraint} synthetic and real-world multivariate time series \newtext{(see SI Section S2.3, Fig. S5; SI section S4, Fig. S8-S10 for more details)}. Note how the distributions of the three real-world datasets employed exhibit noticeable differences in their profile, yet always statistically distinct from the corresponding null models. \newtext{See SI Section S5 and SI Fig.~S11 for a similar analysis with edge-based indicators.}  (\textbf{b}) Two-dimensional histogram of the different contributions associated with 1D cycles in the landscape of coherent and decoherent co-fluctuations. Here, the position of each point in the triangle is determined by the three different contributions associated with the 1D cycles. For example, a point would be at the centre of the triangle if the hyper complexity indicator splits into three equal contributions of Full Coherence (FC), Coherence Transition (CT), and Full Decoherence (FD), while a corner position is reserved for points whose mainly contributions come either from FC, CT, or FD.} 
            \label{fig:fig3_real_world_distributions}
        \end{center}
    \end{figure*}
    \begin{figure*}[t!]
        \centering
        \includegraphics[width=1\textwidth]{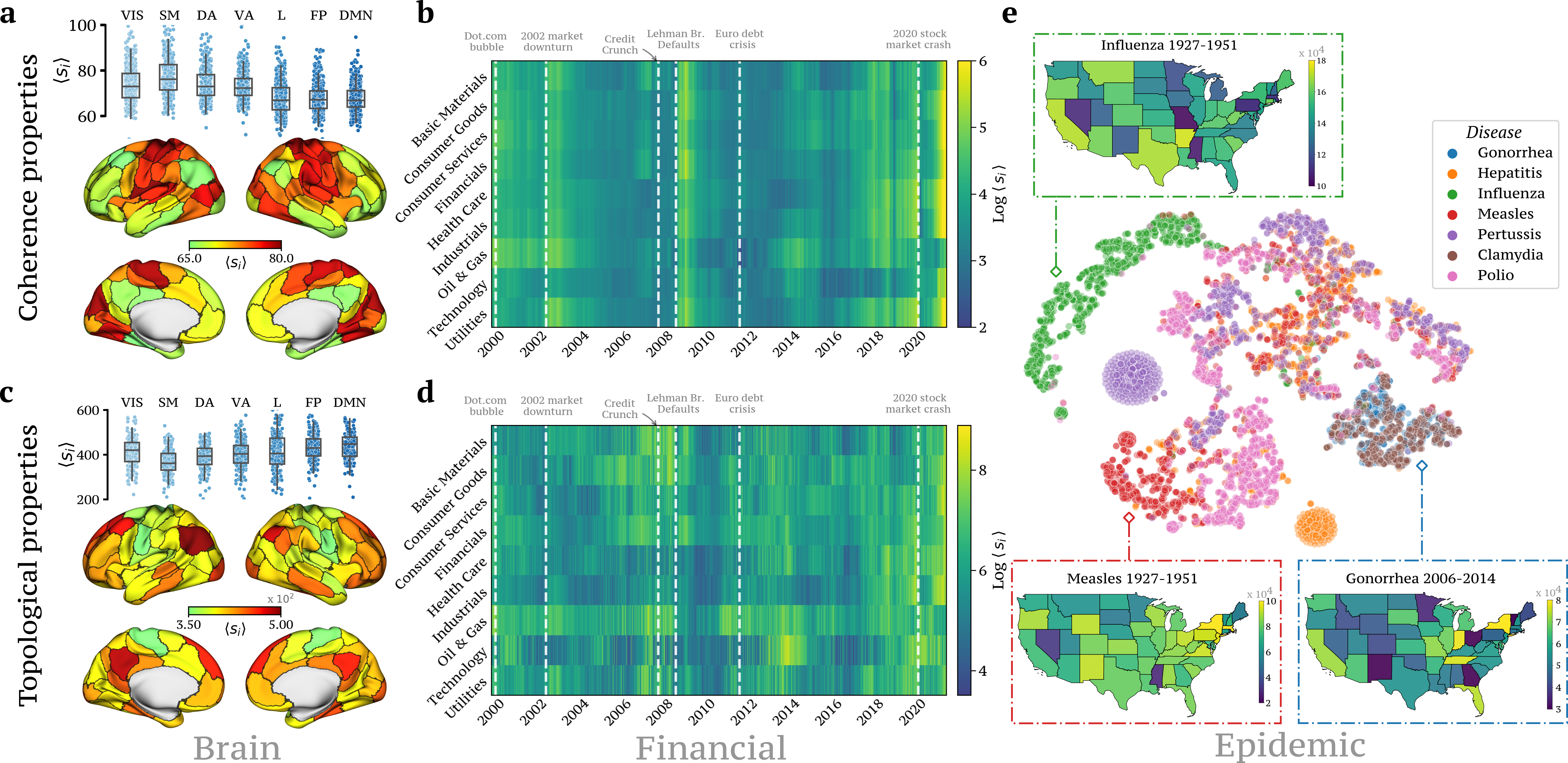}
        \caption{\textbf{Projection of higher-order measures provides local \newtext{spatiotemporal} information}. The nodal strength extracted from the violating triangles $\Delta_v$ can be used to track the importance of higher-order structures in time. (\textbf{a}) Brain map of the nodes involved in higher-order co-fluctuations obtained when isolating the top $15\%$ coherent frames, which are the ones associated with a more synchronized dynamics during rest. \newtext{We also show the boxplots reporting the mean coherence within the Yeo seven canonical functional networks~\cite{yeo2011organization}, namely, the Visual (VIS), SomatoMotor (SM), Dorsal-Attention (DA), Ventral-Attention (VA), Limbic (L), FrontoParietal (FP), and Default Mode Network (DMN).} (\textbf{b}) The temporal evolution of the nodal strength of violating triangles $\Delta_v$ at the level of industrial sectors discriminates crises from periods of financial stability.  (\textbf{c}) The brain map obtained when selecting the 15\% low-hyper complexity frames mainly encompasses the Default Mode Network, \newtext{as confirmed by its high mean nodal score.}  (\textbf{d}) The temporal evolution of the nodal strength of the homological scaffold provides finer details on the downturns of certain economic sectors. Note that, from a topological perspective, the nodal strength extracted from the homological scaffold provides information about 1D loops in the space of co-fluctuations. (\textbf{e}) Finally, a planar embedding of the historical data of epidemic outbreaks can be obtained through t-SNE nonlinear dimensionality reduction when considering as features  the higher-order indicators, such as hyper coherence, the three contributes of hyper complexity, and the average edge violation (see Method for definition). As inset plots, we report the nodal strength of the violating triangles $\Delta_v$  at the US-state level when selecting the $15\%$ high-coherent frames. Remarkably, the spatiotemporal evolution of the outbreaks is different across states and diseases.}
        \label{fig:fig4_local_results}
    \end{figure*}

    In Fig.~\ref{fig:fig3_real_world_distributions}a we report the distributions of hyper coherence for the five CMLs dynamical regimes and the three datasets. For comparison, we also plot the null models obtained by independently reshuffling synthetic and real-world multivariate time series \newtext{(see SI Sections S4 and SI Figures~S8-S10 for the behaviour of the higher-order indicators in more conservative null models)}. 
    Several things can be observed when examining the hyper coherence distributions for real-world systems. First, these distributions are always statistically distinct from the corresponding null models \newtext{(all $p$-values $< 10^{-10}$, with the Kolmogorov-Smirnov test)}, yet they also exhibit specific profiles which strongly differ from each other. If we focus on the epidemic data, for example, it is already possible to differentiate the diseases by coarsely comparing the corresponding hyper coherence distributions. These distributions, in fact, reflect the unique higher-order spatiotemporal patterns inherent to the evolution of the disease. For the financial system, by contrast, we obtain a bimodal distribution mirroring the dichotomy between financial periods of crisis and stability. That is, economic crises are typically characterized by increased (hyper) synchronization, whereas periods of financial stability seem to unfold in a more chaotic fashion. Moreover, armed with the CMLs interpretational benchmarks, we find that, during rest, the human brain is mostly associated with chaotic states and few partially synchronized states, in agreement with studies on resting state brain dynamics~\cite{deco2020turbulentlike,deco2017dynamics,deco2013resting,chialvo2010emergent}.\\
    
    \noindent
    \textbf{Hyper complexity decomposition provides detailed information about dynamical regimes.}
    To better characterize the evolution of 1D homological generators in the space of coherent and decoherent co-fluctuations, we decompose the hyper complexity indicator into three different contributions. That is, as we track the evolution of 1D cycles along the filtration, we focus on 1D cycles that are created and closed only by fully coherent structures, i.e. edges and triangles having a weight larger than zero, which we \newtext{denote} as a Full Coherence (FC) contribution; 1D cycles formed by coherent structures and closed by the decoherent ones (i.e. edges and triangles with a weight smaller than zero), which we \newtext{denote} as a Coherence Transition (CT) contribution;  \newtext{finally,} 1D cycles created only by the fully decoherent structures, we \newtext{denote} as a Full Decoherence (FD) contribution. \newtext{Clearly, by construction, the sum of these three contributions sums up to the total hyper complexity.} We show an illustrative example in SI Fig.~S1.
    
    In Fig.~\ref{fig:fig3_real_world_distributions}b we plot the three fractional contributions to the hyper complexity indicator in a triangular representation.
    In this space, a point is placed on the bottom-left corner if all the 1D cycles are formed and closed only by fully coherent structures. Likewise, the bottom-right corner corresponds to an exclusive contribution from fully decoherent structures, and the top corner corresponds to a contribution uniquely determined by the coherence transition. Whenever the hyper coherence indicator splits into similar FC and FD contributions, the point is placed between the corresponding corners, so that its position reflects the relative importance of the contributions. For example, a point would be at the centre of the triangle if the hyper complexity indicator is split into three equal contributions of FC-CT-FD. Note that such decomposition carries completely different information with respect to the hyper coherence indicator, and yet we draw similar analogies to the results just presented.

    Indeed, when examining the different contributions of the hyper complexity indicator in synthetic signals, we find that the five \newtext{CML} regimes appear to be separated in different clusters. 
    Partially synchronized signals are characterized by a mixture of Fully Coherence and Coherence Transition contributions, while chaotic signals are mainly determined by Fully Coherence and Fully Decoherence (see Fig.~\ref{fig:fig3_real_world_distributions}b left panel). 
    In comparison, for the human brain at rest, we find that most of the states are positioned between chaotic and partially synchronized regimes. This is in agreement with the results obtained when considering the hyper coherence indicator, which provides information of different nature, i.e. it is only based on the number of simplicial violations.
    
    \smallskip
    \textbf{Real-world applications of higher-order topological markers}
    
    So far, we have mainly focused on the temporal evolution of our global higher-order indicators in synthetic and real-world multivariate time series. In what follows, we report some representative applications when considering higher-order measures on a more local level.
    Our goal is to characterize the higher-order states with the largest level of synchronization in both resting-state brain data and financial systems. 
    To this end, in the context of the human brain, we isolated the top $15\%$ coherent frames, which are those associated with a more synchronized dynamical phase. In Fig.~\ref{fig:fig4_local_results}a we report a brain map of the most discriminative nodes by projecting the magnitudes of the violating triangles $\Delta_v$ on a nodal level (see Methods for details and \newtext{SI Fig.~S15} for comparisons at other peaks percentages). This is equivalent to considering the nodal strength extracted from the list of
    violating triangles $\Delta_v$. In other words, regions with the highest absolute value are the ones belonging to the most coherent higher-order structures. In particular, we find activity patterns with emphasized synchronized co-fluctuations mainly reflect sensorimotor areas, which belong to one of the well-known substrates present in the resting-state network~\cite{smith2009correspondence}. \newtext{This is confirmed when considering the histogram reporting the mean coherence within the seven canonical functional networks~\cite{yeo2011organization} (see also SI Fig.~S16 for the effects of higher-order indicators between the functional networks).}
    
    In Fig.~\ref{fig:fig4_local_results}b we report the temporal evolution of the nodal strength extracted from the list of violating triangles $\Delta_v$, yet aggregated at the level of industrial sectors, for the financial time series. The highest values capture the onset of the major periods of financial instability (2002, corresponding to the market downturn, and 2007–2008, corresponding to the great recession that took place as a consequence of the subprime mortgage crisis), which are characterized by an increased synchronization of stock prices, which clearly distinguishes them from the unsynchronized intervals 2002–2007 and 2013-2018, which in turn corresponds to a more stable period of the economy.
    
    Similar analyses can be produced by focusing on the hyper complexity indicator and the nodal strength of the homological scaffold constructed from the persistent homology generators of $H_1$ (see Methods for details). In particular, Fig.~\ref{fig:fig4_local_results}c depicts the brain map obtained when isolating the $15\%$ low-hyper complexity frames, which, as previously shown, are the ones associated with a more synchronized dynamical phase. Here, the highest absolute values are the ones associated with the Default Mode Network (DMN), which is known for being the most active network during wakeful rest~\cite{raichle2001default,fox2005human}.
    
    By contrast, for the financial time series in Fig.~\ref{fig:fig4_local_results}d, the temporal evolution of the nodal strength of the homological scaffold provides fine details on the downturns of certain economic sectors. For instance, consumer goods, basic materials, as well as oil and gas, are the main sectors affected by the great recession of 2007.
    
    Finally, by analysing the historical data of epidemic outbreaks in the US, we show that the temporal evolution of the higher-order measures (i.e. hyper coherence, the three contributions of hyper complexity, and the average edge violations; see Methods for definition) can be used to classify different infectious diseases. In particular, a support vector machine (SVM) classifier reports a high accuracy level, i.e. around 85 \%, using a 10-fold cross-validation setting repeated 50 times (for a comparison between classifiers see SI \newtext{Table~S2}). To provide a more intuitive representation of this result, we report in Fig.~\ref{fig:fig4_local_results}e a planar embedding of the historical data of epidemic outbreaks obtained using the \newtext{t-distributed Stochastic Neighbor Embedding (t-SNE)} nonlinear dimensionality reduction method. Note that nonlinear methods, such as t-SNE, allow to preserve the ``local'' structure in the original high-dimensional space after projection into the low-dimensional space, which is typically not possible with linear methods like \newtext{Principal Component Analysis (PCA) or Multidimensional Scaling (MDS)}~\cite{vandermaaten2008visualizing}.  In this space, we observe that diseases of different kinds cluster together to a great extent, somehow reflecting the unique spatiotemporal evolution of the outbreaks, which are indeed captured by the SVM classifier. At the same time, similarities between diseases can be observed. This is the case for sexually transmitted diseases, such as gonorrhea and chlamydia, which are mostly overlapping in the planar embedding.  As inset plots, we also report the map at the US-state level obtained when selecting the $15\%$ high-coherent frames and considering the nodal strength of the violating triangles $\Delta_v$. We find that the spatiotemporal evolution of the outbreaks is different across states and diseases, somehow reflecting the unique ``higher-order'' characteristics of the disease.

    \tocless{\section*{Discussion}}
    Inferring the dynamics of higher-order structures in multivariate time series is of utmost importance in many complex systems, from epidemiological, to financial, to biological systems. However, direct higher-order network measurements are often inaccessible~\cite{battiston2021physics}.
    As a matter of fact, the vast majority of complex spatiotemporal activity patterns commonly found in many biological, social, and financial systems are typically recorded on a nodal level, rather than directly measured at the level of edges or groups. The higher-order approach introduced in this work provides the first powerful and alternative method to dynamically reconstruct higher-order interactions from multivariate time series.
    
    As a starting benchmark, we have first validated our method against signals whose underlying dynamics is well known. In particular, \newtext{differently from various lower-order statistics~\cite{esfahlani2020highamplitude,macmahon2015community}, the global higher-order indicators presented in this work are able to robustly separate several dynamical phases in high-dimensional coupled chaotic maps, which appear to be distinguishable only through methods based on higher-order statistics~\cite{rosas2019quantifying} (see Fig.~3 and SI Figures~S6, S7). This provides further empirical evidence on the need of higher-order approaches for identifying higher-order behaviours~\cite{rosas2022disentangling}}. 
    On a more local level, i.e.  when projecting the list of hyper coherent triangles as a weighted graph, the graph weight distribution reflects the global dynamic of the multivariate time series: synchronized periodic series convert into well-peaked distributions, while chaotic series convert into fat-tailed distributions. Armed with these theoretical foundations, we then applied our framework to real-world multivariate time series, specifically resting-state fMRI signals from 100 unrelated human subjects, prices of financial assets in the New York Stock Exchange, and historical data from 7 different epidemic outbreaks. 
    
    We found that, during rest, the human brain higher-order dynamics mainly oscillates between fully developed turbulence and partial synchronization. This is in agreement with recent studies supporting that the human brain operates in a turbulent regime~\cite{deco2020turbulentlike,deco2021leonardo}, at the edge of criticality~\cite{perl2022edge}, which seems to confer significant information processing advantages~\cite{deco2021rare}. Moreover, when analysing brain states on a finer scale, we found two notable aspects. On the one hand, the maximally coherent higher-order structures reflect sensorimotor areas, which belong to one of the well-known substrates present in the resting-state brain network architecture~\cite{smith2009correspondence}. These regions are known to play a major role in deciphering --- over very fast time scales~\cite{vandeville2021when} --- inputs that constantly change due to the external environment~\cite{deco2017dynamics}. On the other hand, when examining the hyper complexity marker at its lowest points, we found that the nodal projection coarsely captures the Default-Mode Network (DMN), which is known to integrate high and low-order information in human brain networks~\cite{raichle2001default,fox2005human}. Hence, our two proposed markers provide complementary insights, \newtext{which are not trivially deducible from an edge-wise approach (see SI Section S5 and SI Figures S12, S13),}  on how the brain network segregates and integrates higher-order information over time: hyper synchronous, less integrated, as measured by the number of simplicial violations, for sensorimotor regions; more integrated within the system, as measured in terms of ``proper'' higher-order simplices, for the DMN, whose brain dynamics acts mainly inward, in a constant state of internal exploration through the integration of low and higher-order dynamics~\cite{vandeville2021when,amico2019centralized}, at the ``edge of instability''~\cite{deco2013resting}.
    
    In the context of financial time series, instead, we provided evidence that the magnitude of higher-order structures efficiently discriminate crises from periods of financial stability\newtext{, which cannot be obtained from different null models (see SI Section S4.2).} In particular, maximally coherent higher-order structures strongly emerge in correspondence of major financial crises, mirroring the increase of synchronous co-activation patterns (i.e. stock prices tend to move to the same direction, therefore increasing their level of synchronization). While this is not new in the literature~\cite{peron2011collective,lacasa2015network,kutner2019econophysicsb}, we stress that,  unlike our method, most of these approaches rely on correlation matrices estimated over sliding time-windows~\cite{mantegna1999introduction}, therefore neglecting the information that one might want to capture at the level of individual frames (e.g. in high-frequency trading~\cite{musciotto2021highfrequency}). When examining the hyper complexity indicator at its lowest points, topological markers capture refined information regarding the different role of industrial sectors during crises, revealing strong variations in time and high heterogeneity across different industries~\cite{musmeci2017multiplex},  suggesting their potential to identify the building up of systemic risk~\cite{squartini2013earlywarning,iori2018chapter}.
    
    Finally, when analysing historical data of epidemic outbreaks in the US, we have shown that the temporal evolution of our higher-order measures can be used to classify, to a great extent, diseases of different kind.  In particular, a planar embedding of diseases revealed the presence of interesting clusters based on their unique spatiotemporal pattern. While this result is interesting per se across disciplines~\cite{bishop2006pattern,xing2010brief,kobak2019art,peng2021neural}, our higher-order markers may provide new tools for the quest of epidemic outbreak predictability~\cite{box2015time,pei2018forecasting,scarpino2019predictability}, despite its limitation when increasing forecast length~\cite{scarpino2019predictability,farmer1987predicting}.
    
    Taken together, here we have developed a new flexible tool to provide framewise estimates (therefore circumventing the limitations of sliding window approaches~\cite{hutchison2013dynamic,hindriks2016can}) of higher-order structures in multivariate time-series. 
    We believe that our framework can be effectively used in all situations where the dynamics of signals is poorly understood or unknown, paving the way to further applications in the fields of biology, fluid dynamics, social sciences, or clinical neuroscience. \newtext{In particular, the higher-order indicators and the corresponding lower-order projections can also provide topological \textit{Polaroids}, that is, instantaneous topological snapshots of the spatial configuration of the system under study.}  
    Overall, our approach suggests that investigating the higher-order structure of multivariate time series might provide new insights compared to standard methods, allowing to better characterize group dependencies inherent to real-world data.
    
    \tocless{\section*{Methods}}
    
    \paragraph*{Higher-order topology of multivariate time series ---}
    Let us consider a $N$-dimensional real valued time series $\{\textbf{x}(t)\}_{i=1}^{N}$ with $T$ time points, where the generic time series $\textbf{x}_i=[x_i(1),x_i(2), \ldots, x_i(T)]$ is usually measured empirically or extracted from a $N$-dimensional  deterministic/stochastic dynamical system. It is well established that it is possible to construct correlation matrices by estimating the statistical dependency between every pair of time series~\cite{wei2005time,sporns2010networks}. Here, the magnitude of that dependency is usually interpreted as a measure of how strongly (or weakly) those two time series are related to each other. Following the edge-centric approach proposed in Ref.~\cite{faskowitz2020edgecentric}, however, it is possible to estimate the instantaneous co-fluctuation magnitude between a pair of time series $\textbf{x}_i$ and $\textbf{x}_j$ $-$once they have been $z$-scored$-$ by estimating their element-wise product. That is, for every pair of time series, a new time series encodes the magnitude of co-fluctuation between those signals resolved at every moment in time. We generalise such a concept to the case of higher-order interactions, i.e. triangles, tetrahedron, etc.  We first $z$-score each original time series $\textbf{x}_i$, such that $\textbf{z}_i= \frac{\textbf{x}_i - \mu[\textbf{x}_i]}{\sigma[\textbf{x}_i]}$, where $\mu[\bullet]$ and $\sigma[\bullet]$ are the time-averaged mean and standard deviation. We can then calculate the generic element at time $t$ of the $z$-scored $k$-order co-fluctuations between $(k+1)$ time series as 
    \begin{equation*}
        \xi_{0 \ldots k}(t) =\displaystyle \frac{\prod_{p=0}^{k} z_p(t) - \mu \left[\prod_{p=0}^{k} \mathbf{z}_p \right]}{\sigma \left[\prod_{p=0}^{k} \mathbf{z}_p\right]},
    \end{equation*}
    where also in this case $\mu[\bullet]$ and $\sigma[\bullet]$ are the time-averaged mean and SD functions.  In order to differentiate concordant group interactions from discordant ones in a $k$-order product, concordant signs are always positively mapped, while discordant signs are negatively mapped. Formally, 
    \begin{equation*}
        \mathrm{sign}\left[ \xi_{0 \ldots k} (t) \right] := (-1)^{\mathrm{sgn}\left[(k+1) - \left| \sum_{0}^{k} \mathrm{sgn} \left[ z_i(t) \right]\right| \right]},
    \end{equation*}
    where $\mathrm{sgn}[\bullet]$ is the signum function of a real number. In other words, the weight $w_{0\ldots k}(t)$ at time $t$ of the $k$-order co-fluctuations is defined as:
    \begin{equation*}
        w_{0\ldots k}(t)= \mathrm{sign}[\xi_{0 \ldots k}(t)] \vert \xi_{0\ldots k}(t) \vert
    \end{equation*}
    If we compute all the possible products up to order $k$, this will result in $\binom{N}{k}$ different co-fluctuation time series for each order $k$.
    
    For each time $t$, we condense all the different $k$-order co-fluctuations into a weighted simplicial complex $\mathcal{K}^t$.  Formally, a $(d-1)$-dimensional simplex $\sigma$ is defined as the set of $d$ vertices, i.e. $\sigma= [p_0,p_1,\ldots,p_{d-1}]$. A collection of simplices is a simplicial complex $\mathcal{K}$ if for each simplex $\sigma$ all its possible subfaces (defined as subsets of $\sigma$) are themselves contained in $\mathcal{K}$~\cite{hatcher2005algebraic}. Weighted simplicial complexes  are simplicial complexes with assigned values (called weights) on the simplices.
    
    For simplicity, in this work we only consider co-fluctuations of dimension up to $k=2$, so that triangles represent the only higher-order structures in the weighted simplicial complex $\mathcal{K}$, and weights on the simplices, i.e. $w_{ij}$ and $w_{ijk}$, represent the magnitude of edges and triangles co-fluctuations. 
    
    \newtext{Note finally that, in order to compare our approach with the edge-based approach, we employed the Root Sum Square (RSS) of the edge-time series Ref.~\cite{esfahlani2020highamplitude}, which can be used as a direct proxy of the amplitude of the collective co-fluctuations of the edge time series. In other words,  we compute the amplitude of the edge time series as the root sum of squared co-fluctations, i.e. $RSS(t)= \sqrt{\sum_{i,j>i}e_{ij}(t)^2}$, where the vector ${\bf e_{ij}}= {\bf z}_i \,{\bf z}_j$ is the 1-order co-fluctuation (i.e. the edge time series) obtained as a product of the z-scores of the original time series.}

    \paragraph*{Hyper coherence and hyper complexity ---}
    \newtext{To analyse the structure of the weighted simplicial complex $\mathcal{K}^t$ across multiple scales, we  consider a topological data analysis approach~\cite{wasserman2018topological}, which has been shown to unveil new dynamical properties of different complex systems~\cite{lum2013extractinga,nicolau2011topology,saggar2018towards,saggar2022precision}. In particular, we rely on persistent homology, which is a recent technique in computational topology that has been largely used for the analysis of high dimensional datasets~\cite{ghrist2008barcodes,carlsson2008local} and in disparate applications~\cite{lee2011discriminative,carstens2013persistent,horak2009persistent}.} The central idea is the construction of a sequence of successive simplicial complexes, \newtext{which approximates with} increasing precision the original weighted simplicial complex. This sequence of simplicial complexes, i.e. $\emptyset=\mathcal{S}_{0}  \subset \mathcal{S}_1 \subset \ldots \subset \mathcal{S}_l \subset \ldots \subset \mathcal{S}_n$, is such that $S_i \subset S_j$ whenever $i < j$ and is called a filtration.  In our case, we construct a filtration building upon these steps:
    \begin{itemize}
        \item Sort the weights of the links and triangles in a decreasing order: the parameter $\epsilon_l \in \mathbb{R}$ scans the sequence. Equivalently, $\epsilon_l$ is the parameter that keeps track of the actual weight as we gradually scroll the list of weights.
        \item At each step $l$, remove all the triangles that do not satisfy the simplicial closure condition, i.e. $\exists ! (i,j) : w_{ij} < w_{ijk}$. Such triangles are considered as a violation and inserted, along with the corresponding weights, in the \textit{list of violations} $\Delta_v = \{(i,j,k),w_{ijk}\}$. The remaining links and triangles with a weight larger than $\epsilon_l$ belong to the simplicial complex $\mathcal{S}_l$
    \end{itemize}
    We then define the \textit{hyper coherence indicator}, as the fraction of violating coherent triangles (i.e. violating triangles with a weight greater than zero) over all the possible coherent triangles (i.e. triangles with a weight greater than zero). Notice also that when identifying each violating triangle (i.e by checking whether the triangle is entering the complex before its edges), we can keep track of the number of its edges $e_v \in [0,2]$ that are already in the complex. We can then define the \textit{average edge violation} indicator as the total number of those edges $e_v$ averaged over all the violating triangles.
    
    Persistent homology studies the changes of the topological structure along the filtration $\{\mathcal{S}_l\}$ and provides a natural measure of robustness for the topological features emerging across different scales. In particular, it is possible to keep track of these topological changes by looking at each $k$-dimensional cycle in the homology group $H_k$. In our case, we focus on the $1$-dimensional holes (i.e. loops), therefore analysing the homology group $H_1$.  More precisely, at each step of the filtration process, a generator $g$ uniquely identifies a $1$-dimensional cycle by its constituting elements. The importance of the 1-dimensional hole $g$ is encoded in the form of ``time-stamps'' recording its birth $b_g$ and death $d_g$ along the filtration $\{\mathcal{S}_l\}$~\cite{petri2014homological}. These two time-stamps can be combined to define the persistence $\pi_g$ = $d_g -b_g$ of the one-dimensional cycle, which gives a notion of its importance in terms of its lifespan.
    
    A typical way to visualize the results of persistent homology group $H_1$ is through multiset points in the two-dimensional persistence diagram. In this diagram, each point $(b_g, d_g)$ represents a one-dimensional hole $g$ that appears across the filtration. As a consequence, this diagram is a
    compressed summary describing how long 1D cycles live along the filtration and can be used as a proxy of the ``complexity'' of the underlying space. 
    \newtext{In fact, the sum of the persistences of the homological generators of $H_k$ can be seen as the distance of the topological space from the trivial space (i.e. the space without $k+1$-dimensional holes).}
    In our case, we define the \textit{hyper complexity indicator} as the Wasserstein distance~\cite{carriere2017sliced} between the persistence diagram of $H_1$ and the empty persistence diagram, corresponding to a space with trivial $H_1$ homology.
    \newtext{Finally, note that in SI Section S2.2 and SI Fig.~S4, we briefly investigate the presence of 1D cycles and 3D-cavities in the context of CML when extending our framework to 4-body interactions.}

    \paragraph*{Homological scaffold and lower-order projections ---}
    To obtain a finer description of the topological features present in the persistent diagram, we consider the persistence homological scaffold as proposed in Ref.~\cite{petri2014homological}. In a nutshell, this object is a weighted network composed of all the cycle paths corresponding to generators $g_i$ weighted by their persistence $\pi_{g_i}$.  In other words, if an edge $e$ belongs to multiple 1-dimensional cycles $g_0, g_1, \ldots , g_s$, its weight $\bar{w}_e^\pi$ is defined as the sum of the generators' persistence, i.e.:
    $$
    \bar{w}_e^\pi=\sum_{g_i | e \in g_i} \pi_{g_i}
    $$
    The information provided by the homological scaffold allow us to decipher the role that different links have regarding the homological properties of the system. A large total persistence $\bar{w}_e^\pi$ for a link $e$ implies that such link acts as a locally strong bridge in the space of coherent and decoherent co-fluctuations~\cite{petri2014homological}.
    
    Lastly, to analyse the information provided by the list of violations $\Delta_v$ on an edge/node level, we rely on downward projections. That is, for each edge $(i,j)$ we assign a weight $w_{ij}$ equal to the \newtext{average} sum of the weights of triangles defined by that edge, i.e.  triangles of the form $(i,j,\bullet)$ with a weight \newtext{$w_{ij\bullet}$, and the average is computed over the number of triangles $n_{ij\bullet}$ defined by that edge}. Similarly, we define the nodal strength $w_i$ of node $i$ as the \newtext{average} sum of weights of the triangles connected to node $i$.
    In the case of the homological scaffold, since it is a weighted network, the node strength $\bar{w}_i$  of node $i$ is defined, in the classical way~\cite{barrat2004architecture,latora2017book}, as the sum of the weights of edges connected to the node $i$.
    
    \paragraph*{Real-world dataset ---}
    We analysed three datasets belonging to different domains. Specifically, we considered fMRI resting-state data from The Human Connectome Project (HCP, http://www.humanconnectome.org/), the stock prices of the NYSE financial market obtained from the Yahoo! finance API~\cite{yahoodata}, and the historical data of several infectious diseases in the US~\cite{scarpino2019predictability,vanpanhuis2013contagious}. 
    
    The fMRI dataset used in this work consists of resting-state data from 100 unrelated subjects (54 females, 46 males, mean age$ = 29.1 \pm 3.7$ years) as provided at the HCP 900 subjects data release~\citet{vanessen2012human,vanessen2013wuminn}.  We added some extra steps to the HCP minimal preprocessing pipeline~\cite{glasser2013minimal,smith2013restingstate}: First, we applied a standard general linear model regression that included detrending and removal of quadratic trends; removal of motion regressors and their first derivatives; removal of white matter, cerebrospinal fluid signals, and their first derivatives; and global signal regression (and its derivative). Second, we bandpass-filtered the time series in the range of 0.01 to 0.15 Hz. Last, the voxel-wise fMRI time series were averaged into the $N=100$ corresponding brain nodes of the Schaefer cortical atlas~\cite{schaefer2018localglobal} and then z-scored. For completeness, 19 sub-cortical regions were added, as provided by the HCP release~\cite{glasser2013minimal}. The interested reader can refer to Ref.~\cite{vandeville2021when} for details on these steps. 
    
    The financial dataset used in this study was obtained from the Yahoo! finance historical data API (via the Python
    library \textit{yfinance}~\cite{yahoodata}). We have collected the daily prices of 119 US companies in the NYSE from Yahoo! finance in the period from January 1, 2000 to November 30, 2021.
    
    We considered the weekly historical data at the US state-level of several infectious diseases including chlamydia, gonorrhea, influenza, measles, polio, and pertussis. This dataset was previously used in Ref.~\cite{scarpino2019predictability} and is freely available.

    \paragraph*{Limitations ---}
    
    One of the main limitations of our approach concerns the time complexity. Indeed, if we consider co-fluctuation patterns up to the order $k$, the resulting time complexity scales as $\mathcal{O}(N^k)$. Moreover, at the current stage, our framework does not allow to investigate the causality effect between two subsequent time frames (i.e. how much previous time points affect the next ones in terms of the proposed topological markers). \newtext{Notice also that our dynamical higher-order approach, as many of other existing pairwise dynamical methods~\cite{tagliazucchi2012criticality,liu2013timevarying,esfahlani2020highamplitude}, can be heavily affected by noisy fluctuations in the time series. However, this issue can be smoothed out by analyzing statistics averaged over multiple time frames, as we have done in this work. Furthermore, the higher-order brain maps reported in Figure~\ref{fig:fig4_local_results} appear to be robust after accounting for the presence of head motion volumes in the fMRI data (see also SI Fig. 14). Finally, we stress that our framework, with the exception of the hyper complexity indicator, mainly detect coherent synchronous patterns while it mostly ignores the effect of decoherent patterns, which are known to be important in the overall dynamics of a system.  Future work should explore alternative approaches that deal in a more explicit way with decoherent patterns present in the data. }
    \newtext{\paragraph*{Data availability ---}
        The data used for this analysis will be available on Zenodo upon acceptance of the manuscript.
    }
    \paragraph*{Code availability ---}
    The python code used in this work will be made available upon acceptance of the manuscript on AS EPFL webpage, as well as a maintained version on E.A.’s GitHub page (\url{https://github.com/eamico}).

\let\oldaddcontentsline\addcontentsline
\renewcommand{\addcontentsline}[3]{}
\let\addcontentsline\oldaddcontentsline

    \cleardoublepage
    
    \renewcommand\theequation{{S\arabic{equation}}}
    \renewcommand\thetable{{S\arabic{table}}}
    \renewcommand\thefigure{{S\arabic{figure}}}
    \renewcommand\thesection{{S\arabic{section}}}
    \renewcommand{\thesubsection}{{\thesection.\arabic{subsection}}}
    \setcounter{secnumdepth}{2}

    \setcounter{section}{0}
    \setcounter{table}{0}
    \setcounter{figure}{0}
    \setcounter{equation}{0}
    \onecolumngrid

    \begin{center}
\LARGE{Supplementary Material:\\ Unveiling the higher-order organization of multivariate time series}
    \end{center}
\vspace*{0.3 cm}

    \tableofcontents

\cleardoublepage
\section{Filtration and persistent homology}

In the main text, we introduced the simplicial filtration $\mathbb{F}(\mathcal{K}^t)$ as one of the main steps of our higher-order framework. More precisely, for each time $t$, this object represents a sequence of embedded
simplicial complexes, sorted according to coherent
patterns, starting with the empty complex and
ending with the entire simplicial complex $\mathcal{K}^t$, i.e. $\emptyset=\mathcal{S}_{0} \subset \mathcal{S}_1 \subset \ldots \subset \mathcal{S}_l \subset \ldots \subset \mathcal{S}_n \subset \mathcal{K}^t$.  We then considered the persistent homology of $\mathbb{F}$ to characterize the persistency of 1D cycles in
the filtration, i.e. the persistent generators of the
first homology group $H_1$, which provide insights
about where and when higher synchronised regions
emerge.

\begin{figure*}[bh!]
    \centering
    \includegraphics[width=0.95\textwidth]{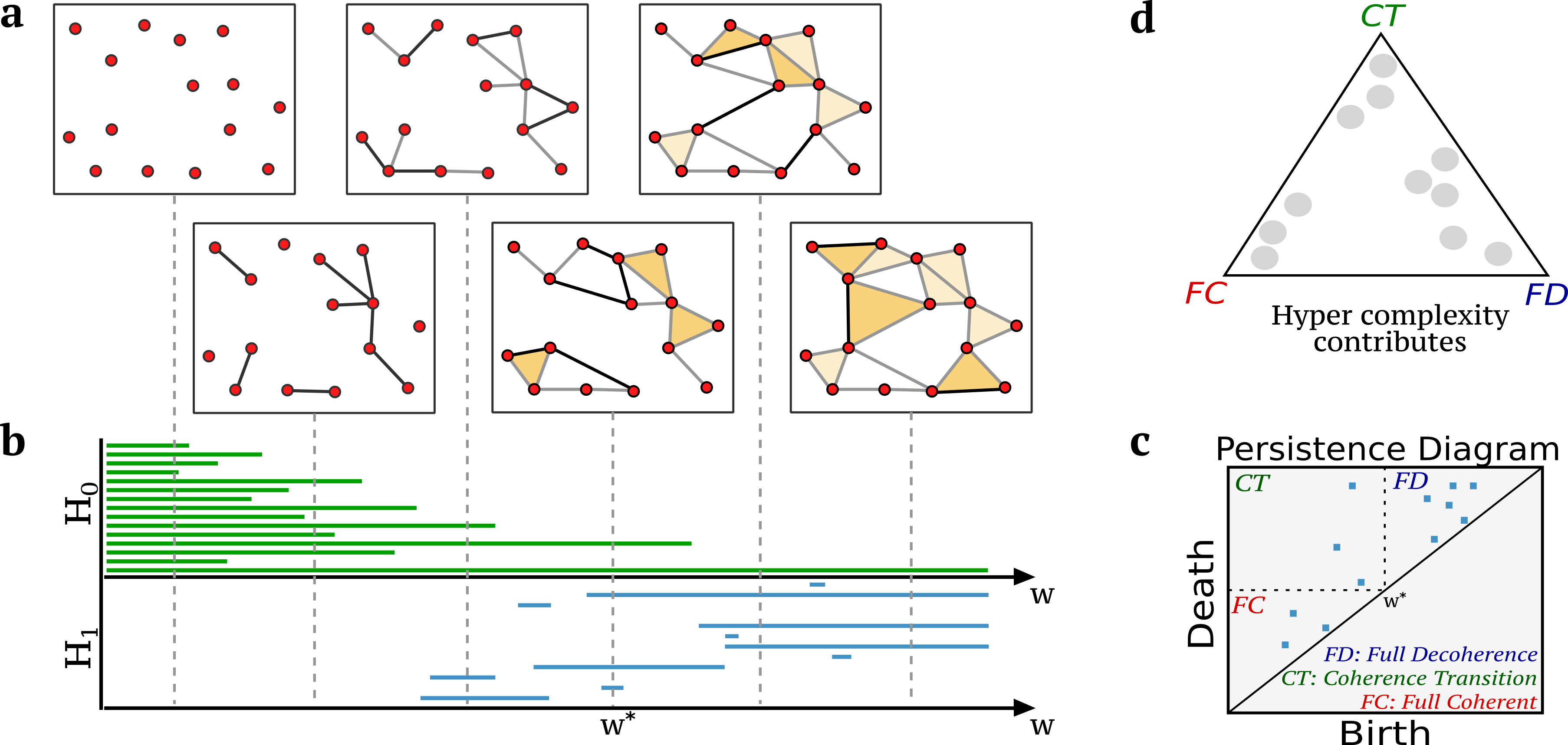}
    \caption{\textbf{Pictorial representation of persistent homology computation and definition of hyper complexity contributes.} \textbf{(a)} Example of a simplicial filtration in two-dimensions. \textbf{(b)} Barcodes describe the lifetime of different topological features across multiple scales.  Here, green bars identify the persistence of various connected components (describing $H_0$), progressively merging into each other until only one survives, whereas blue bars describe the lifetime of 1-dimensional cycles (describing $H_1$). Hence, each bar corresponds to a specific topological feature, which can be identified in terms of a ``time-stamp'' recording the birth $w_b$ and death $w_d$ of that feature along the filtration. \textbf{(c)} Persistence diagrams provide an equivalent description of barcodes. For example, if we focus only on $H_1$, each 1D cycle is represented in the 2-dimensional plot by a point with coordinates $(w_b,w_d)$. In this work, we further distinguish the nature of 1D cycles depending on the corresponding time-stamps, i.e. whether 1D cycles are created and closed before/after $w^*=0$, therefore reflecting pure coherent (resp. incoherent) structures. \textbf{(d)} The hyper complexity indicator, defined as the Wasserstein distance~\cite{carriere2017sliced_SI} between the persistence diagram of $H_1$ and the empty persistence diagram, can be then decomposed into three different contributes according to the nature of 1D cycles and plotted in a triangular representation.}
    \label{fig:SI_fig1_PH}
\end{figure*}

In Fig.~\ref{fig:SI_fig1_PH} we report a schematic representation of the persistent homology computation and the corresponding definition of the hyper complexity contributes considered in this work.  In particular,  Fig.~\ref{fig:SI_fig1_PH}a reports an illustrative example of a simplicial filtration in two-dimensions.
The outputs of persistent homology are barcodes, which represent a compressed description of the homological features of a space (see Fig.~\ref{fig:SI_fig1_PH}b). Each bar corresponds to a specific topological feature, which can be identified in terms of a ``time-stamp'' recording the birth $w_b$ and death $w_d$ of that feature along the filtration. In the main text, we characterized the evolution of 1D cycles along such a filtration --- the blue bars in Fig.~\ref{fig:SI_fig1_PH}b --- by relying on persistence diagrams, which provide an equivalent description of bar codes. In this 2-dimensional plot, each 1D cycle is represented by a point with coordinates $(w_b,w_d)$ (see  Fig.~\ref{fig:SI_fig1_PH}c).

Furthermore, we defined the \textit{hyper complexity indicator} as the Wasserstein distance~\cite{carriere2017sliced_SI} between the persistence diagram of $H_1$ and the empty persistence diagram, corresponding to a space with trivial $H_1$ homology. However, to better characterize the evolution of 1D cycles in the space of coherent and decoherent co-fluctuations, we decompose the hyper complexity indicator into three different contributions. That is, as we track the evolution of 1D cycles along the filtration, we focus on 1D cycles that are created and closed only by fully coherent structures, i.e. edges and triangles having a weight larger than zero, which we renamed as a Full Coherence (FC) contribution; 1D cycles formed by coherent structures and closed by the decoherent ones (i.e. edges and triangles with a weight smaller than zero), which we renamed as a Coherence Transition (CT) contribution;  1D cycles created only by the fully decoherent structures, renamed as a Full Decoherence (FD) contribution.

In Fig.~\ref{fig:SI_fig1_PH}d, the three fractional contributions of the hyper complexity indicator are reported in a triangular representation.

\cleardoublepage
\newpage
\section{Global behaviour of Coupled Map Lattice (CML)}
\begin{figure*}[th!]
    \centering	
    \includegraphics[width=1\textwidth]{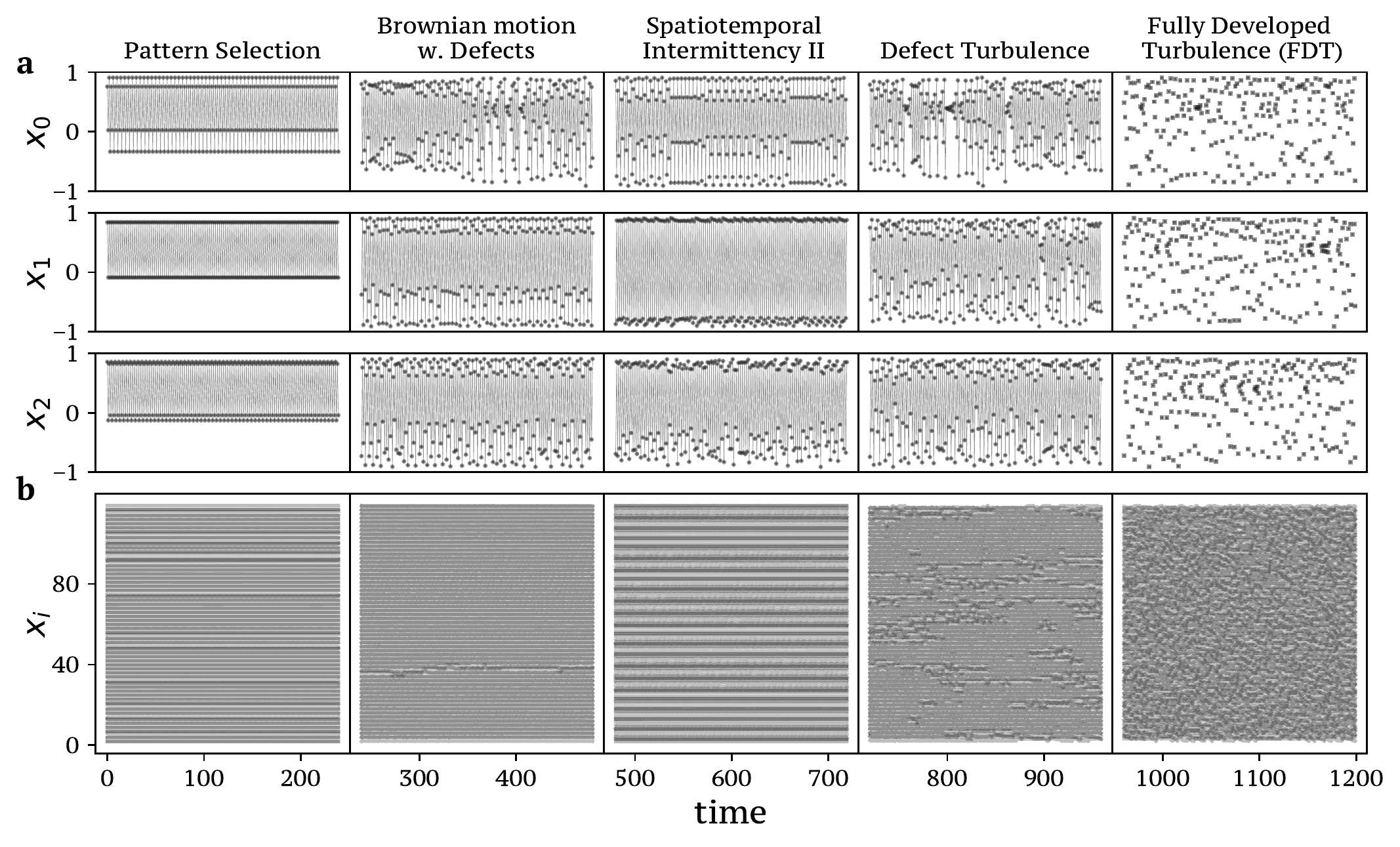}
    \caption{\textbf{Sample of the coupled map lattice} (\textbf{a}) We report a sample of the time series considered in the main text and generated by $N=119$ diffusively coupled fully chaotic maps at different values of the coupling strength $\varepsilon$. \newtext{Namely, from ordered to disordered systems: Pattern Selection (PS) at $\varepsilon= 0.12$, Brownian motion with Defects (BMWD) at $\varepsilon=0.08$, Spatiotemporal Intermittency II (STI) at $\varepsilon = 0.3$, Defect Turbulence (DT) at $\varepsilon = 0.068$, and Fully Developed Turbulence (FDT) at $\varepsilon = 0.05$.} \textbf{(b)} The space-amplitude plot, reporting the temporal evolution of $x_i(t)$ at each lattice site $i$, clearly reflects the different global behaviour of the five dynamical regimes.}
    \label{fig:SI_fig2_coupled_maps}
\end{figure*}

In the main text, we considered $N=119$ diffusively coupled fully chaotic maps to generate synthetic multivariate time series with different dynamical regimes. In Fig.~\ref{fig:SI_fig2_coupled_maps}a we report a sample of such time series in three different lattice sites, showing the general behaviour of the five different dynamical states.  Moreover, the space-amplitude plot~\cite{kaneko1989pattern_SI,kaneko1992overview_SI} reported in Fig.~\ref{fig:SI_fig2_coupled_maps}b, which depicts the temporal evolution of $x_i(t)$ at each lattice site $i$, clearly shows the overall behaviour of the five dynamical regimes.

\begin{figure*}[th!]
    \centering	
    \includegraphics[width=1\textwidth]{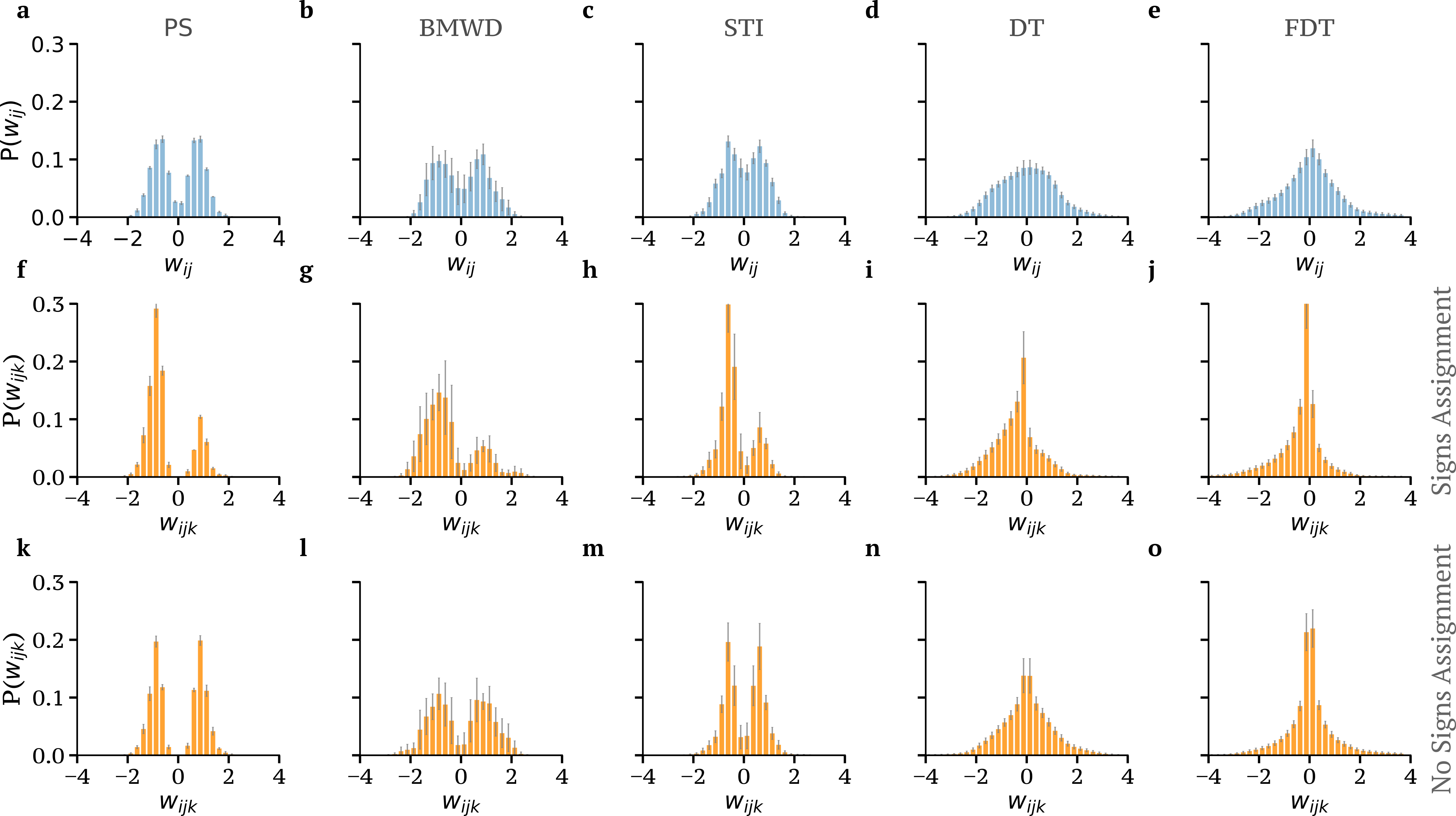}
    \caption{\newtext{\textbf{Impact of $z$-score and signs assignment in edge and triangles distributions for the coupled map lattice.} For each of the five CML dynamical states, we report the distribution of weights of edges \textbf{(a-e)} and triangles with \textbf{(f-j)} and without the sign assignment \textbf{(k-o)}, respectively. Notice that the $z$-scores and signs assignment produce a negative offset for triangles with respect to that of edges.  While this phenomenon does not have a deep impact in the hyper coherence indicator for the way such a measure is constructed, the hyper complexity indicator might be shifted by such offset. It is also worth remarking that when sign remapping is not performed, it is not possible to identify a triangle with a  synchronous co-activation pattern from its weight, given that decoherent triangles might end up having the same weights as fully coherent ones. Error bars represent standard deviations over 240 time points.}}
    \label{fig:SI_fig2b_impact_zscores}
\end{figure*}

\newtext{\subsection{Effect of $z$-scores and signs assignment in CML time series}}
\newtext{In one of the main steps of the simplicial framework described in the main text, we rely on the $z$-score for each group product time series to allow comparison across $k$-orders and to create proper simplicial filtrations. After that,
    a choice on how to assign signs to the resulting weights is required in order to distinguish fully concordant group interactions (all positive or negative fluctuations) from discordant ones (a mixture of positive and negative fluctuations) in a $k$-order product. In order to distinguish fully concordant group interactions (all positive or negative fluctuations) from discordant ones (a mixture of positive and negative fluctuations) in a $k$-order product, we assign positive signs to the fully concordant group interactions, and negative signs to the discordant ones. However, 
    this way of assigning signs means that the baseline score might be slightly altered and, as a result, the concordance might be harder to achieve for a triangle than for an edge.\\
    In Figure~\ref{fig:SI_fig2b_impact_zscores}, we analyse the impact of $z$-score and signs assignment in edges and triangles distributions for the coupled map lattice dynamical states. In particular, the $z$-scores and signs assignment for triangles produce a negative offset with respect to that of edges. This phenomenon does not have a deep impact in the hyper coherence indicator, since it is defined as the fraction of violating coherent triangles (i.e. violating triangles with a weight greater than zero) over all the possible coherent triangles (i.e. triangles with a weight greater than zero). By contrast,  the negative offset of the triangle weight distribution over the edges one might have an impact in the hyper complexity indicator. This is due to the fact that this indicator is constructed by considering the ensemble of both edges and triangles weight distribution. 
    However, the effect on hyper complexity is a shift in the contributions from the fully coherent to the other two contributions, rather than a restructuring of the whole simplicial filtration. In this sense, the effect is a biasing one.}\\

\begin{figure}[h]
    \centering
    \includegraphics[width=0.95\textwidth]{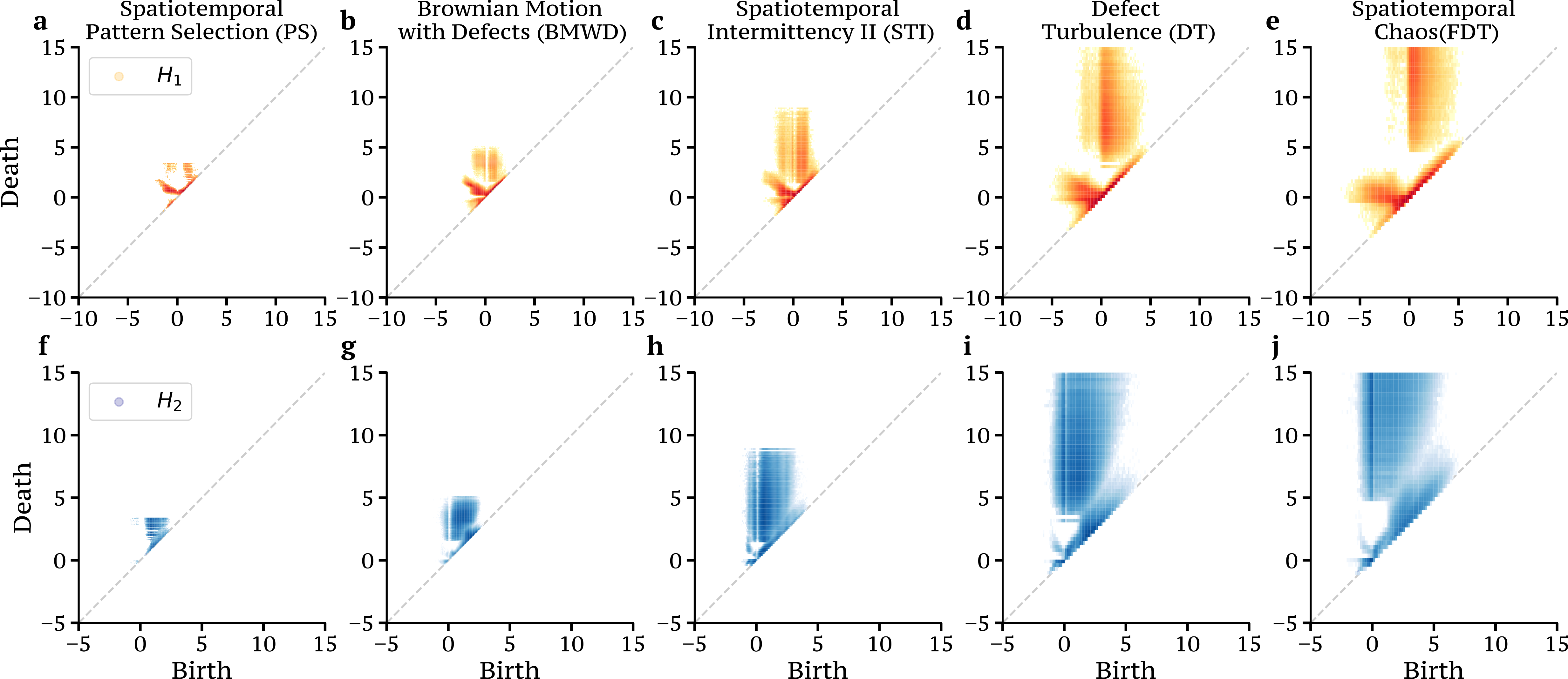}
    \caption{\newtext{\textbf{2D histograms for the $H_1$ and $H_2$ generators in Coupled Map Lattices}. We report, for each dynamical state of the Coupled Map Lattice with $N=50$, the persistence diagram for the homology groups $H_1$ and $H_2$. Interestingly, the  $H_2$ persistence diagram is different between the CML dynamical regimes and, as a result, it might help to further differentiate the five dynamical states when considered in synergy with $H_1$. Results are obtained considering $10$ independent realization and $240$ time points for each block in analogy with the results reported in the main text.  The dynamical states are: Pattern Selection (PS) at $\varepsilon= 0.12$, Brownian motion with Defects (BMWD) at $\varepsilon=0.08$, Spatiotemporal Intermittency II (STI) at $\varepsilon = 0.3$, Defect Turbulence (DT) at $\varepsilon = 0.068$, and Fully Developed Turbulence (FDT) at $\varepsilon = 0.05$.}}
    \label{fig:SI_fig4_cavities}
\end{figure}

\newtext{\subsection{Impact of 3D-cavities}}
\newtext{In this section we briefly investigate the presence of 3D-cavities (i.e. the generators of the homology group $H_2$)  when including in the analysis all the interactions up to 4-body interactions.  Indeed, in the main text, we mainly focused on analyzing 1D cycles (i.e. the generators of the homology group $H_1$) of the simplicial filtration $\mathbb{F}(\mathcal{K}^t)$ that we generate at each time $t$, while we explicitly did not consider 3D-cavities. This is because we limited our analysis up to 3-body interactions (i.e. up to triangles), and 3D-cavities do not add meaningful information. As a matter of fact, if only 3-body interactions are present, then all the 3D-cavities have an infinite persistence due to the absence of 4-body interactions that close cavities. In other words, without 4-body interactions, 3D-cavities will have a birth in the filtration but not a death. Nevertheless, the presence of cavities in the simplicial filtration $\mathbb{F}(\mathcal{K}^t)$ might add important information when characterizing the higher-order structure of a multivariate time series, as shown when analysing the human connectome data~\cite{sizemore2018cliques_SI}.\\
    Figure~\ref{fig:SI_fig4_cavities} reports the persistence diagrams for the homology groups $H_1$, $H_2$ obtained in the context of Coupled Map Lattices with $N=50$ nodes when extending our framework to 4-body interactions. Remarkably, we find that the persistence diagram for $H_2$ is diverse for the five dynamical regimes and, as a results, it might be used together with $H_1$ to further differentiate the CML dynamical regimes.}

\begin{figure}
    \centering
    \includegraphics[width=0.6\textwidth]{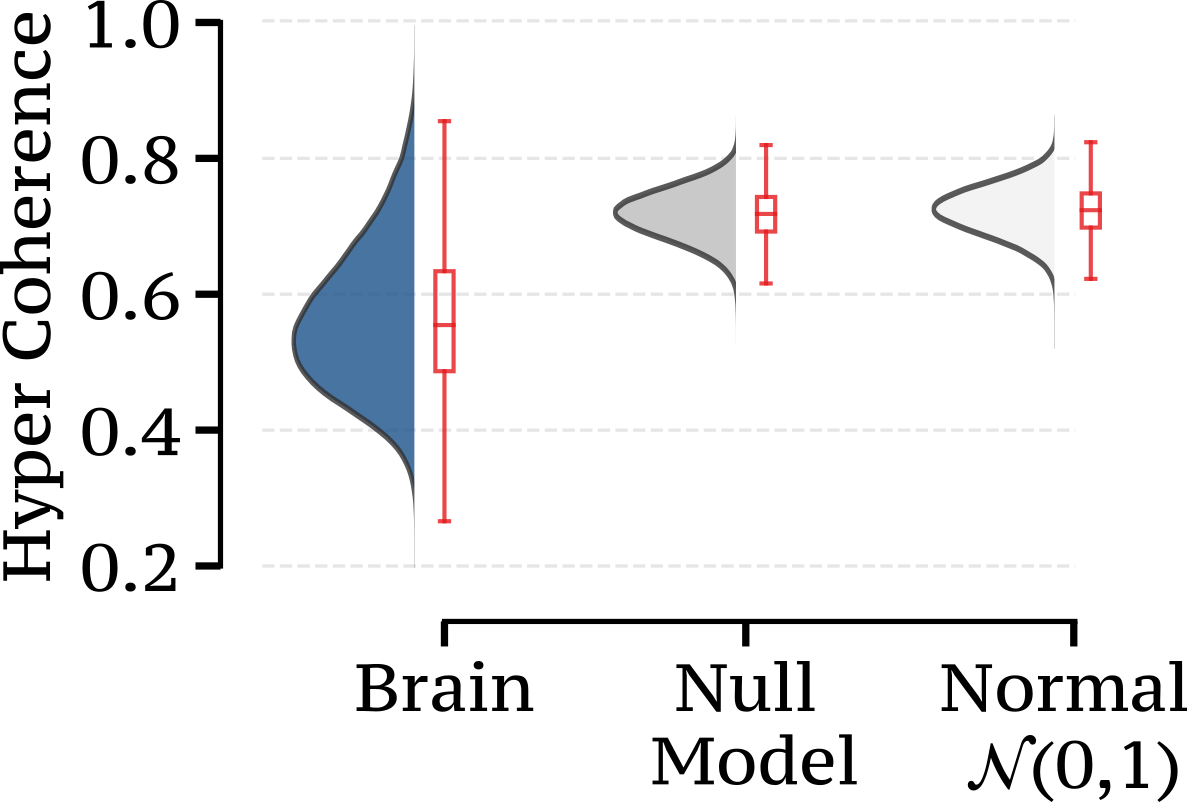}
    \caption{\newtext{\textbf{Hyper coherence distribution computed from resting-state fMRI data and Gaussian multivariate time series}. Violin plots showing the distribution of Hyper Coherence for the resting-state fMRI data (N=119 brain regions), the corresponding null model obtained when independently reshuffling the time series, and for a multivariate time series, where each of the time series is sampled from a normal distribution $\mathcal{N}(0,1)$. The null model of the resting-state fMRI data leads to a distribution of Hyper Coherence that is statistically equivalent to the one obtained when analysing Gaussian multivariate time series ($p$-value $< 10^{-10}$ with the Kolmogorov-Smirnov test).}}
    \label{fig:SI_fig4_gaussian}
\end{figure}

\newtext{\subsection{Higher-order approach on Gaussian multivariate time series}}
\newtext{In this section we show the Hyper Coherence distribution obtained when considering a $N$-dimensional Gaussian multivariate time series, where each of the time series is sampled from a normal distribution $\mathcal{N}(0,1)$ with zero mean and unit variance.\\
    We report in Figure~\ref{fig:SI_fig4_gaussian} the violin plots reporting the hyper coherence distribution computed from resting-state fMRI data, the corresponding null model obtained when independently reshuffling the time series, and from Gaussian multivariate time series. It is interesting to notice how the null model of the resting-state fMRI leads to a hyper coherence distribution that is statistically equivalent to the one obtained when analysing Gaussian multivariate time series ($p$-value $< 10^{-10}$ with the Kolmogorov-Smirnov test). This is due to the fact that, even though each BOLD signal is not Gaussian per se, the entire ensemble can be considered as Gaussian. As a result,  the simple shuffling procedure destroys all the temporal dependencies and the resulting co-fluctuations of the multivariate signals tend to a Gaussian distribution.}

\cleardoublepage
\newpage

\newtext{\section{Comparison of time series approaches to distinguish the CML regimes}}
In this section, \newtext{we provide a detailed investigation of} the ability of different indicators to differentiate the five regimes generated by $N = 119$ diffusively coupled fully chaotic maps, extensively considered in the main text. The multivariate time series consists of $N=119$ nodes and  $T=1200$, obtained by concatenating five different CML regimes with fixed time length  $L=240$.\\
\newtext{We first compare our higher-order measures with the lower-order dynamical indicator originally proposed in Ref.~\cite{esfahlani2020highamplitude_SI}. This is the Root Sum Square (RSS) of the edge-time series, which was recently used to identify important ``events'' in fMRI signals~\cite{esfahlani2020highamplitude_SI,pope2021modular_SI}, and it is a direct proxy of the amplitude of the collective co-fluctuations of the edge time series. In other words,  we compute the amplitude of the edge time series as the root sum of squared co-fluctations, i.e. $RSS(t)= \sqrt{\sum_{i,j>i}e_{ij}(t)^2}$. Here, the vector ${\bf e_{ij}}= {\bf z}_i \,{\bf z}_j$ is the edge time series obtained as a product of the $z$-scores of the original time series. We then assess the performance of several other ``static'' approaches, namely, \textit{(i)} the information-theoretic approaches introduced in Refs.~\cite{rosas2019quantifying_SI,gatica2021highorder_SI} accounting for higher-order interactions,  \textit{(ii)} the dyadic method at the interface of network science and random matrix theory~\cite{macmahon2015community_SI}, and \textit{(iii)} the classical approach based on Pearson's correlation coefficient~\cite{sporns2010networks_SI}.} 

\begin{figure*}[b!]
    \centering
    \includegraphics[width=1\textwidth]{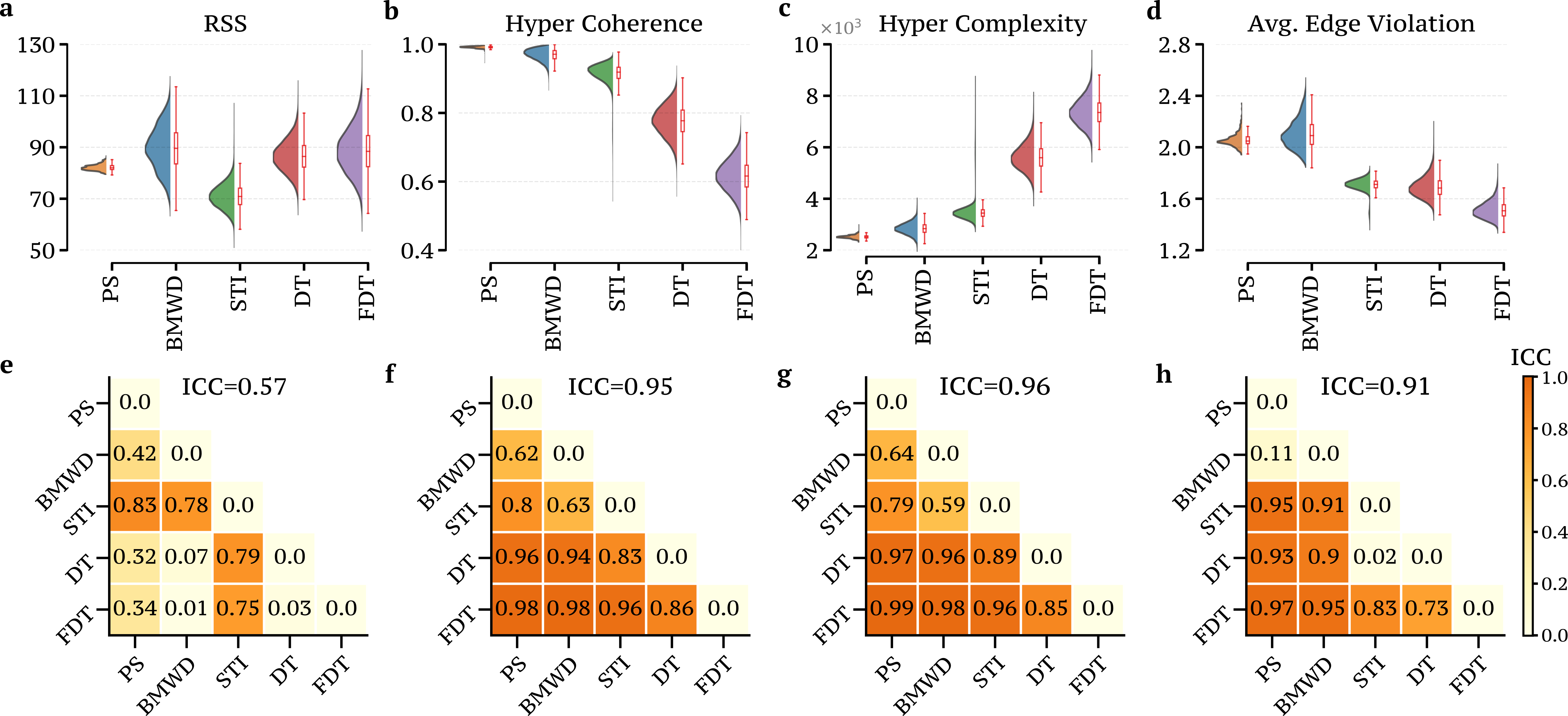}
    \caption{\newtext{\textbf{Comparison of different dynamical metrics}. \textbf{(a-d)} Violin plots showing the distribution of RSS (Root Sum Square) and three higher-order measures for the five different dynamical regimes generated by the diffusively coupled chaotic maps. Remarkably, the RSS statistic, which only captures the effect of lower-order structures (i.e. the edges), is not able to well-separate the five dynamical regimes. By contrast, the higher-order metrics introduced in this work distinguish the different regimes. \textbf{(e-h)} The performance of each indicator is quantitatively measured by the $ICC$ values computed either considering the five distributions or between all the possible pairs. When considering all the five distributions, we find RSS with an $ICC \approx 0.57$,  hyper coherence with $ICC \approx 0.95$, hyper complexity with $ICC \approx 0.96$, and average edge violation with $ICC \approx 0.91$. Results are averaged over 100 independent realizations.}}
    \label{fig:SI_fig3_metric_comparison}
\end{figure*}

\newtext{Top panels of Figure~\ref{fig:SI_fig3_metric_comparison} depict the violin plots of the distribution of RSS and of the three higher-order measures for the five regimes of the CMLs.} Surprisingly, the RSS distributions of certain dynamical states (BMWD, DT, and FDT) are highly similar to each other, somehow mirroring the inability to capture the subtleties of these regimes. By contrast, the higher-order measures seem to differentiate the dynamical states in a qualitatively better way. As a matter of fact, the ability to distinguish the different dynamical regimes is quantitatively confirmed by the $ICC$ values associated with each metric \newtext{and reported in the bottom panels. In particular, we have RSS with an $ICC \approx 0.57$,  hyper coherence with $ICC \approx 0.95$, hyper complexity with $ICC \approx 0.96$, and average edge violation with $ICC \approx 0.91$. For the sake of completeness, for each dynamical measure we also report an ICC matrix encoding the ICC values between all the possible pairwise comparison of dynamical regimes.}

\begin{figure*}[b!]
    \centering
    \includegraphics[width=0.98\textwidth]{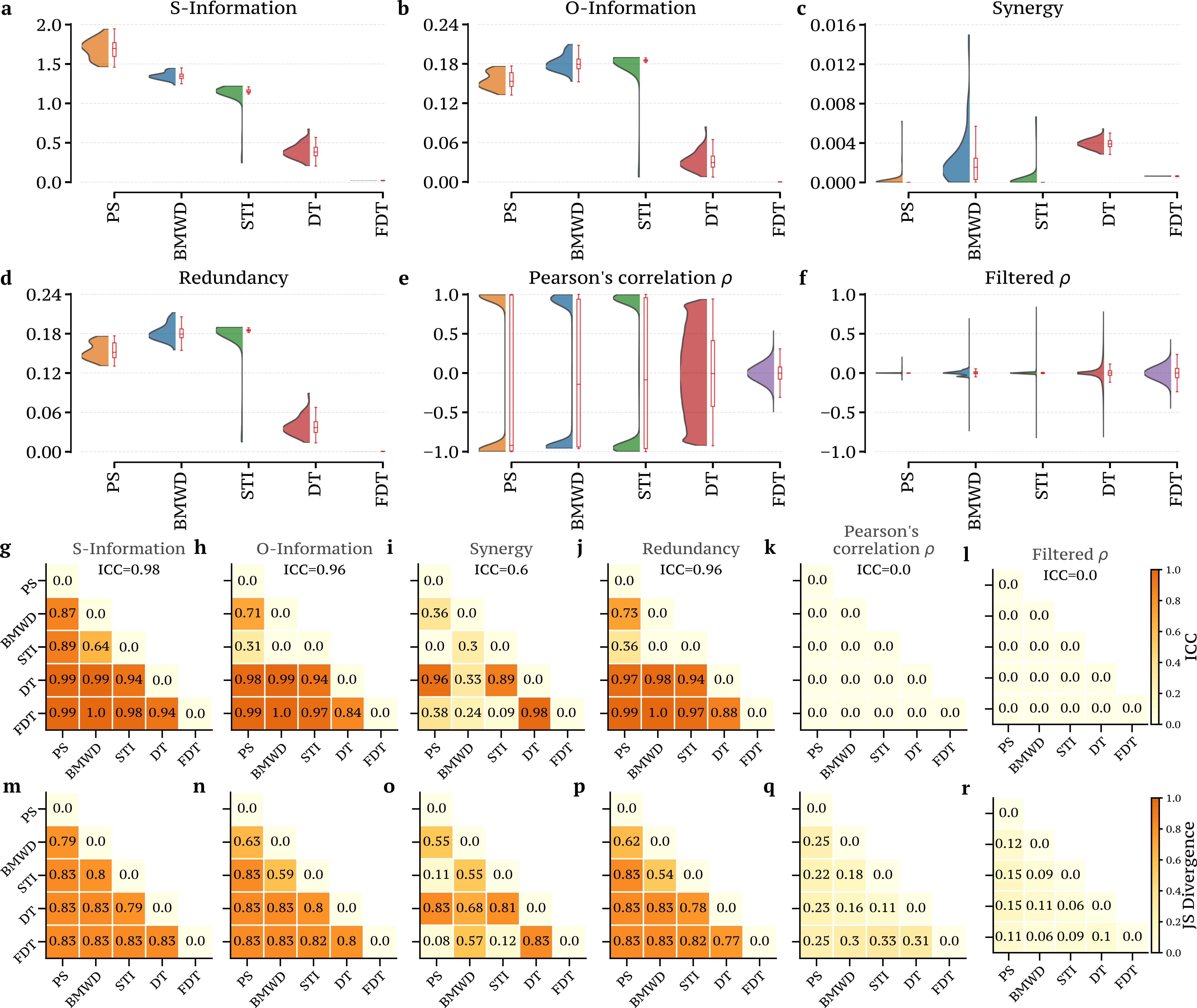}
    \caption{\newtext{\textbf{Comparison between approaches for distinguishing the CML regimes}. \textbf{(a-f)} For the five dynamical regimes generated by the diffusively coupled chaotic maps, we report the violin plots showing the distribution of ``static'' higher-order measures~\cite{rosas2019quantifying_SI}, and ``static'' pairwise measures, namely, the Pearson's correlation coefficient $\rho$ and filtered $\rho$ obtained by considering the approach based on random matrix theory~\cite{macmahon2015community_SI}. Remarkably, only higher-order measures are able to separate the five dynamical regimes. \textbf{(g-l)} We report the $ICC$ values associated with each metric, either computed considering the five distributions or between all the possible pairs, thus demonstrating that higher-order measures outperform statistics based on pairwise correlations. \textbf{(m-r)} We repeat the same analysis but considering the Jensen-Shannon (JS) divergence to measure distances between distributions, leading to the same conclusions of the ICC approach.}}
    \label{fig:SI_fig4_metric_comparison_static}
\end{figure*} 
\newtext{Lastly, for each of the five CML regimes, we independently compute diverse ``static'' higher-order and pairwise measures~\cite{sporns2010networks_SI,macmahon2015community_SI,rosas2019quantifying_SI,gatica2021highorder_SI}, aggregating the results across 100 independent realizations.  In other words, after isolating each dynamical state by selecting the precise intervals of size $T=240$, we compute the selected measure on that interval. This is necessary, since such measures are not dynamic (i.e. they are not defined on a single frame), but instead they require a time-window to be computed.}

\newtext{
    Figure~\ref{fig:SI_fig4_metric_comparison_static} reports the results of our analyses for the different approaches. Remarkably, several of the information-theoretic measures introduced by Rosas and colleagues~\cite{rosas2019quantifying_SI} [see Figs.~\ref{fig:SI_fig4_metric_comparison_static}(a-d)] are able to differentiate in a quantitative way the different CML dynamical states, as confirmed by the high ICC values reported in panels~\ref{fig:SI_fig4_metric_comparison_static}(g-j)\footnote{\newtext{The code for these analyses has been adapted from Ref.~\cite{gatica2021highorder_SI}.}}. By contrast, approaches only based on pairwise statistical dependencies [see Figs.~\ref{fig:SI_fig4_metric_comparison_static}(e-f)], such as ``connectome'' analyses based on Pearson's correlation $\rho$ or the filtered $\rho$ relying on random matrix theory~\cite{macmahon2015community_SI},  fail at differentiating the five dynamical regimes [see Figs.~\ref{fig:SI_fig4_metric_comparison_static}(k-l)]. To confirm the quantitative analyses provided by the ICC metric, we repeat an analogous analysis but considering the Jensen-Shannon (JS) divergence to measure distances between distributions [see Figs.~\ref{fig:SI_fig4_metric_comparison_static}(m-r)]. }

\cleardoublepage
\newpage
\newtext{\section{Impact of different null models}}
\newtext{\subsection{CML synthetic multivariate time series}}
\newtext{After having demonstrated that standard metrics to assess two-body similarity of time series, such as RSS, pearson correlation, or more refined approaches based on random matrix theory~\cite{macmahon2015community_SI}, fail to distinguish the CML dynamical regimes, we now shift the focus to investigating the impact of alternative null models~\cite{schreiber2000surrogate_SI,donges2015unified_SI,vavsa2022null_SI}. Indeed, the null reported in Figs.~2-3 of the main text --- obtained by independently shuffling the time series --- might be in principle too simplistic to detect regime-specific features of the co-fluctuations in the CML case. This is because the empirical signal of the co-fluctuations in each dynamical regime is compared to products of node-level signals at different times and for different regimes. As a consequence, we test the global behaviour of our two higher-order indicators when considering two alternative null models, namely, the \textit{(i)} block null model, and the (\textit{ii)} phase randomization null model.\\
    The first null model, hereafter denoted as \textit{block null model}, is constructed such that the shuffling of each dynamical regimes of the CML is performed separately for each region (i.e. the blocks consisting of 240 time points). In this way,  we are ensuring that the dynamical regimes are not mixed during the reshuffling, and we can therefore detect regime-specific features of the co-fluctuations.\\
    The second null model, hereafter denoted as \textit{phase randomization null model}, is a conventional approach~\cite{schreiber2000surrogate_SI} to generate surrogate time series preserving the empirical power spectra, while randomizing the temporal dependencies. 
    More specifically, such time series are generated by transforming the CML time series to the frequency domain via Fourier transform, shuffling the phase coefficients, and then taking the inverse transform to the time domain~\cite{vavsa2022null_SI}.}

\begin{figure*}[h!]
    \centering
    \includegraphics[width=0.98 \textwidth]{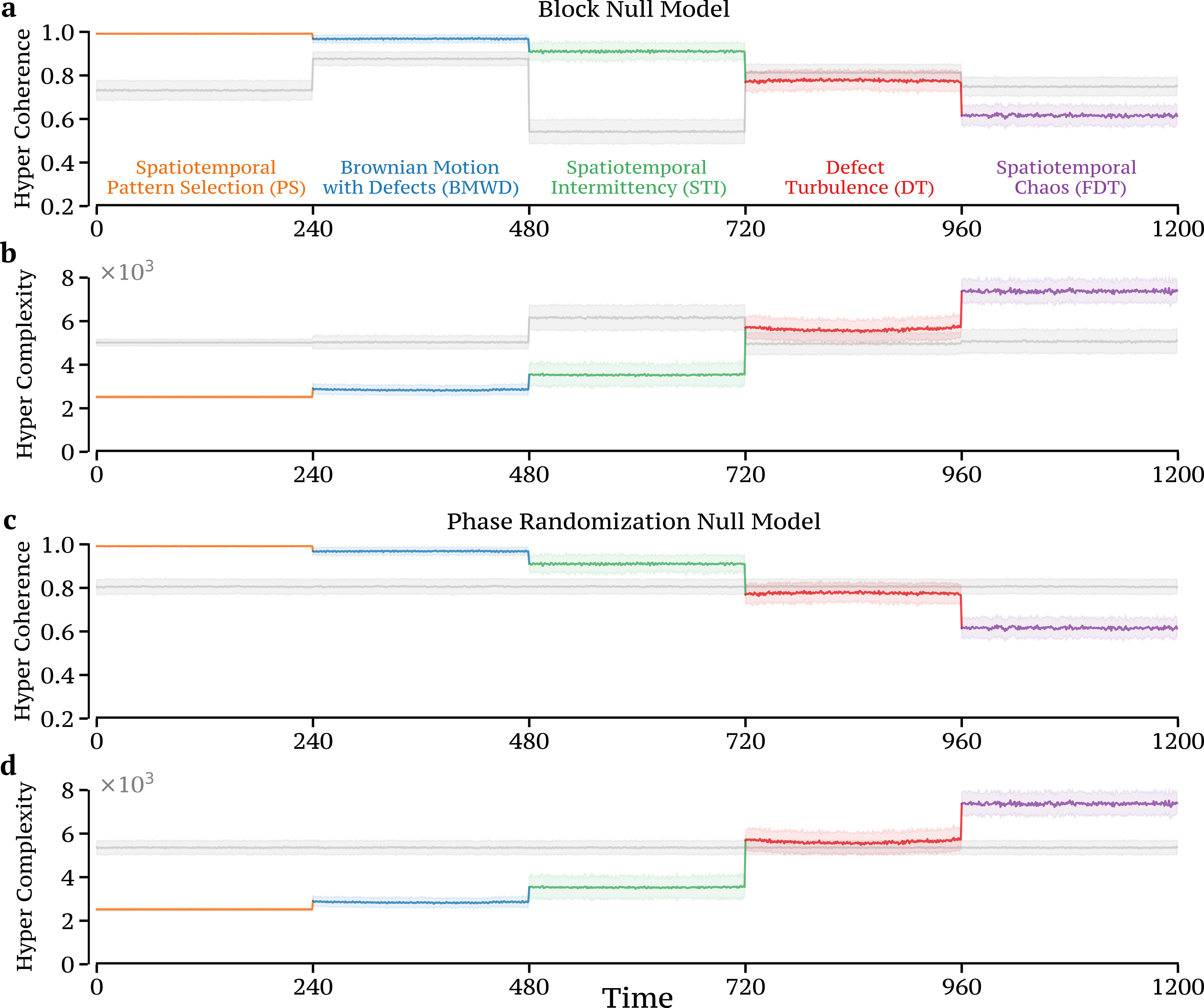}
    \caption{\newtext{\textbf{Global behaviour of our higher-order indicators for two alternative null models.} \textbf{(a-b)} We report the behaviour of the two higher-order indicators, as in Fig.~2 of the main text, respectively computed for the original CML and for the surrogate time series generated by the block null model, which shuffles independently each block of 240 time points. Remarkably,  the  resulting blocks of hyper coherence in this null model become less distinguishable from each other, while keeping certain regime-specific features of the co-fluctuations. However, the ranking from ordered to disordered states is now lost. In addition, the blocks of hyper complexity cannot be separated in the block null model, somehow supporting the intuition that this topological-based measure is more fine-grained than the hyper coherence. \textbf{(c-d)} The same analyses are repeated when considering the phase randomization null model, which preserves the power spectra of the CML time series.  } }
    \label{fig:SI_fig5_global_behaviour_nullmodels}
\end{figure*}

\newtext{We first aim at replicating the panels of Fig.~2(a,c) of the main text when considering the two alternative null models. In particular, Figure~\ref{fig:SI_fig5_global_behaviour_nullmodels} summarizes the results of our higher-order approach when applied to the two surrogate multivariate series. We find that the distributions of hyper coherence in the block null model become less distinguishable from each other, while keeping certain regime-specific features of the co-fluctuations. Furthermore, the block null model does not preserve any more the ranking between order (i.e. the PS regime) and disorder (i.e. the Spatiotemporal Chaos), see Fig.~\ref{fig:SI_fig5_global_behaviour_nullmodels}(a). In addition, the distributions of hyper complexity in the block null model are mostly identical to each other for the five blocks, so that it is clear that there are some intrinsic topological properties that are only present in the CML multivariate time series [see Fig.~\ref{fig:SI_fig5_global_behaviour_nullmodels}(b)].  When considering the phase randomization null model, we find a flat behaviour for both the hyper coherence and hyper complexity  distributions, see Fig.~\ref{fig:SI_fig5_global_behaviour_nullmodels}(c-d).}

\begin{figure*}[h!]
    \centering
    \includegraphics[width=0.98 \textwidth]{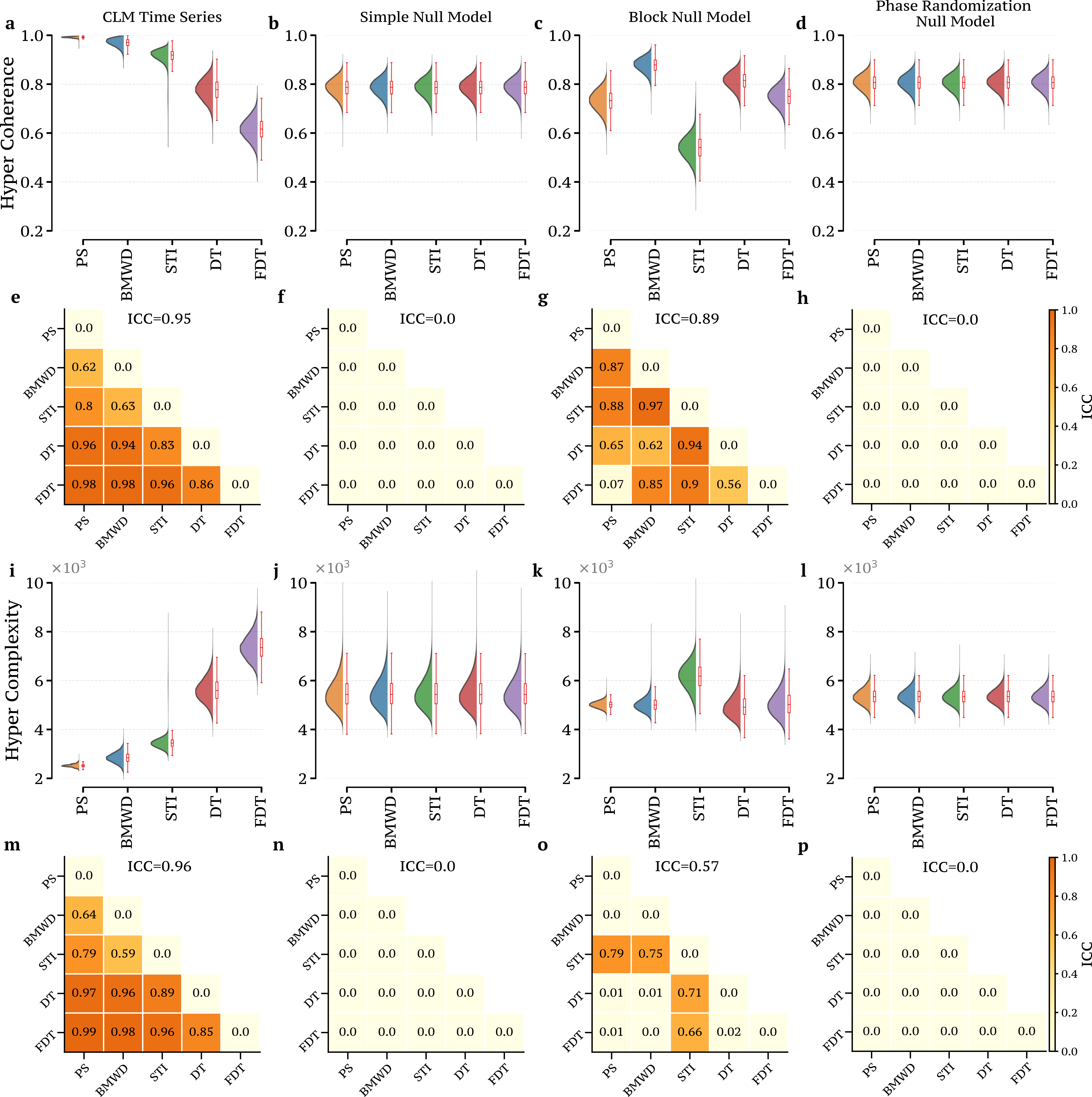}
    \caption{\newtext{\textbf{Comparison of null models in Coupled Map Lattice.} \textbf{(a-d)} We report the distribution of hyper coherence for the classical CML time series and for the three null models considered in this work, namely, the simple null model considered in the main text, the block null model, and the phase randomization null model. \textbf{(e-h)} Both the global and pairwise analysis of the ICC metric reveal that only the block null model is able to partly separate the CML regimes, hence keeping certain regime-specific features of the co-fluctuations, while breaking the order-disorder gradient present in the real Hyper Coherence distributions. \textbf{(i-l)} When repeating the analyses of panels (a-d) for the distribution of hyper complexity, we find that the distributions of hyper complexity in the null models are mostly identical to each other for the five blocks, so that it is clear that there are some intrinsic topological properties that are only present in the CML multivariate time series. \textbf{(m-p)} Global and pairwise analysis of the ICC metric provide a quantitative way of assessing the similarities between each regimes in the different null models.}}
    \label{fig:SI_fig6_quantitative_comparisons_nullmodels}
\end{figure*}

\newtext{As done in previous sections, to quantitatively test the ability to distinguish the different CML dynamical states in the null models,  we rely again on the ICC values, which are either computed considering the five distributions or between all the possible pairs of regimes. We report in Figure~\ref{fig:SI_fig6_quantitative_comparisons_nullmodels} the distributions of hyper coherence and hyper complexity for the actual CML time series and for all the different surrogates time series.}

\cleardoublepage
\newpage
\newtext{\subsection{Block null model in financial time series}}
\newtext{Here, we assess whether the results of hyper coherence of the financial time series presented in the main text can be trivially obtained by analysing block null models which account for different global trends, including upwards and downwards ones. 
    As a matter of fact, in the considered time span of 21 years (2000-2021) the market experienced different global trends. Therefore, when considering a simple null model, as the one described in the main text,  which independently reshuffles each stock price in the whole window, we might neglect the importance of different financial trends. In order to assess this potential confound, we consider three different block null models with respectively 3, 6, and 9 temporal blocks, and test whether the findings on financial time series in the main text could not be simply deduced from more sophisticated null models. The blocks are selected in order to capture the major financial trends of the last 21 years and to consider crises as midpoints of equally spaced time windows.}

\begin{figure}[hb!]
\centering
\includegraphics[width=1\textwidth]{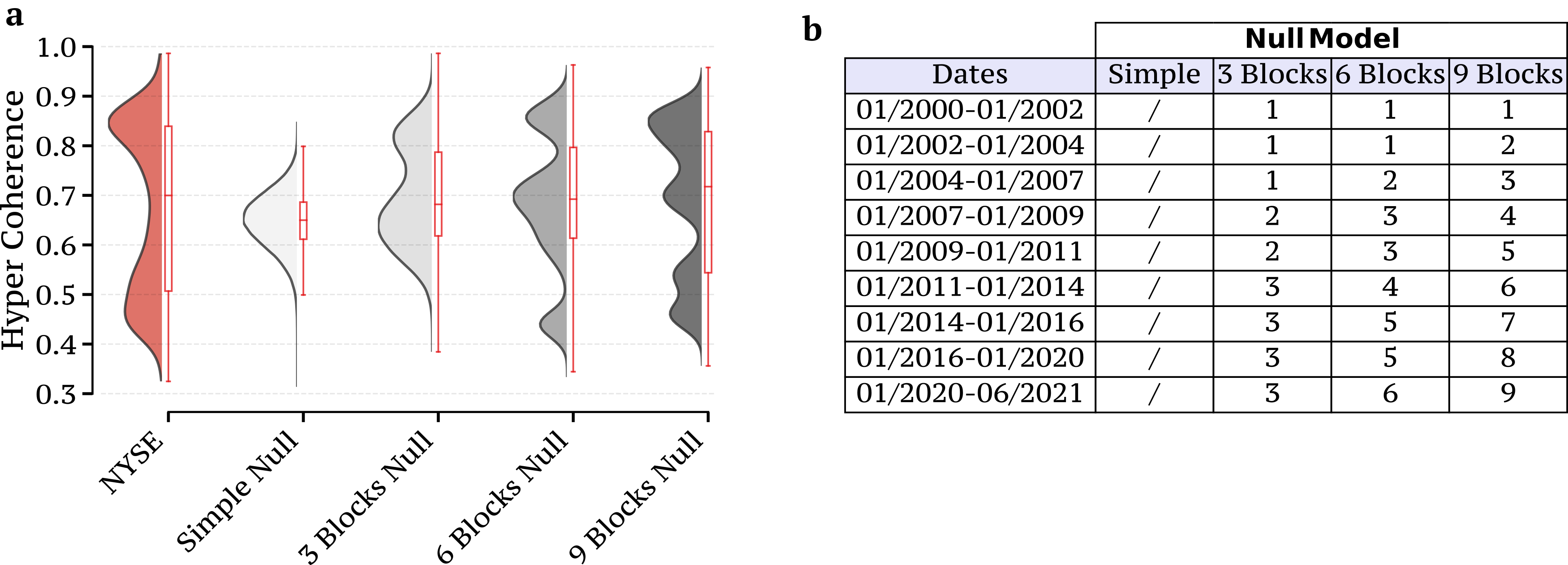}
\caption{\newtext{\textbf{Distribution of hyper coherence for different block null models}. 
        \textbf{(a)} Violin plots showing the distribution of
        hyper coherence for the financial prices of 119 assets in NYSE spanning a period from 01/2000 to 06/2021. We also report the corresponding null models obtained when independently reshuffling the empirical time series over the whole time window of 21 years (i.e. Simple Null), or restricting the shuffling over 3, 6, or 9 blocks, respectively. Even when considering the null model obtained with 9 blocks (dark grey curve), the hyper coherence distribution is distinct from the one obtained for the empirical multivariate time series (red curve).
        \textbf{(b)} We report the periods considered for the various null models, which include finer information about the global financial trends. For instance, the 3 block null model spans three different periods, namely, $01/2000-01/2007$, $01/2007-01/2011$, and $01/2011-06/2021$.}}
\label{fig:SI_fig8_blocknulls_financial}
\end{figure}

\newtext{In Figure~\ref{fig:SI_fig8_blocknulls_financial} we report the distributions of hyper coherence for the financial dataset along with the four null models obtained by independently reshuffling the real-world multivariate time series over the whole time window of 21 years, or restricting the shuffling over 3, 6, or 9 temporal blocks, respectively. It is interesting to notice that null models with finer information about the global financial trends seem to increasingly approximate the distribution of hyper coherence of the empirical dataset, despite some notable differences are present even in the finer model consisting of 9 temporal blocks. For example, the distribution of the 9 blocks null model is almost 4-modal, while it is bi-modal in the empirical case, and the two distribution are statistically different, i.e. $p<10^{-10}$ after the Kolmogorov-Smirnov test. Finally, for the sake of completeness, we report the list of periods considered for the block null models in Figure~\ref{fig:SI_fig8_blocknulls_financial}(b).}
\cleardoublepage
\newpage
\newtext{\section{Edge-based indicators in real-world systems}
In this section we investigate the impact of edge-based statistics in the context of real-world multivariate time series when compared to the higher-order approach presented in the main text.  More specifically, we first analyse the RSS distributions for the three types of real-world systems considered in this work. Subsequently, we aim at reproducing some of the results of Fig.~5 of the main paper by relying on metrics derived from RSS or, more in general, from measures based on edge signals. Our final goal is to compare several different group-order representations in the context of the real-world systems considered in this study.}\\

\begin{figure}[h!]
\centering
\includegraphics[width=0.95\textwidth]{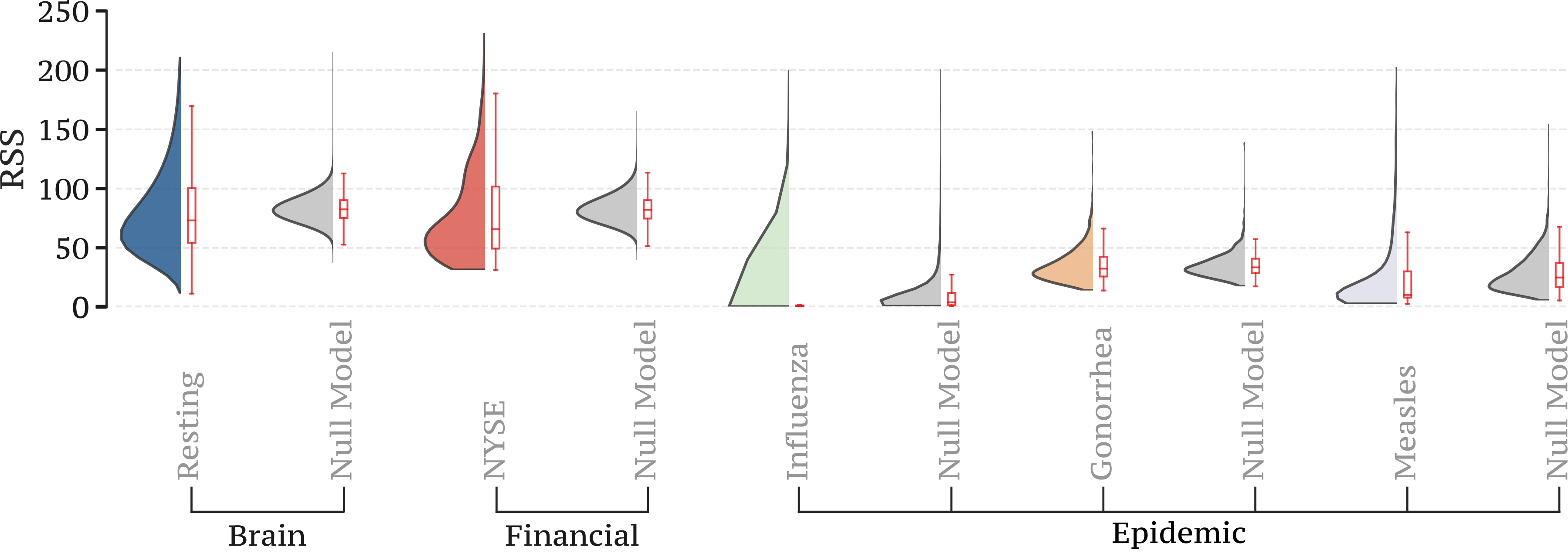}
\caption{\newtext{\textbf{RSS distribution for the empirical time series.} 
        Violin plots showing the distribution of RSS for the three real-world datasets considered in the main text, namely, resting-state fMRI data (N=119 brain regions), financial prices of 119 assets in NYSE in the period 2000-2021, and the US historical data of several infectious diseases at the US state-level (N=50). While for the brain and financial systems the empirical RSS distributions are different from the corresponding null models (obtained by independently reshuffling each time series), the RSS distributions for the epidemic time series are very similar to the ones obtained when considering the corresponding null models ($p <0.001$ for Gonorrhea, with the Kolmogorov-Smirnov test). } }
\label{fig:SI_fig9_RSSdistributions}
\end{figure}

\noindent
\newtext{As global analysis, we report in Figure~\ref{fig:SI_fig9_RSSdistributions} the distributions of RSS for the three real-world datasets. For the sake of comparison, we also plot the null models obtained by independently reshuffling the real-world multivariate time series. We first notice that, in contrast with the results reported in the main text, these distributions are not always statistically distinct from the corresponding null models, as evident when looking at the distributions of the epidemic data (e.g. $p <0.001$ for Gonorrhea with the Kolmogorov-Smirnov test). Moreover, when examining the RSS distribution of resting-state brain and of financial systems, it is not possible to extract insights on the nature of different temporal dynamics present in the system (e.g. we cannot discriminate crises from periods of financial stability).}

\newtext{Once we established that global edge-wise measures such as RSS provide moderate information on the empirical systems, we now shift the focus on investigating the impact of different group-order representations on a more local level. In particular,  we performed several analyses on the empirical time series by projecting the magnitudes of higher- and lower-order approaches on nodal level and compare the overall results. Yet, we remind that any lower-order projection might lead to a moderate/high reduction of the total amount of information and therefore such representations might be potentially misleading if not carefully analysed.}

\begin{figure}[ht!]
\centering
\includegraphics[width=0.85\textwidth]{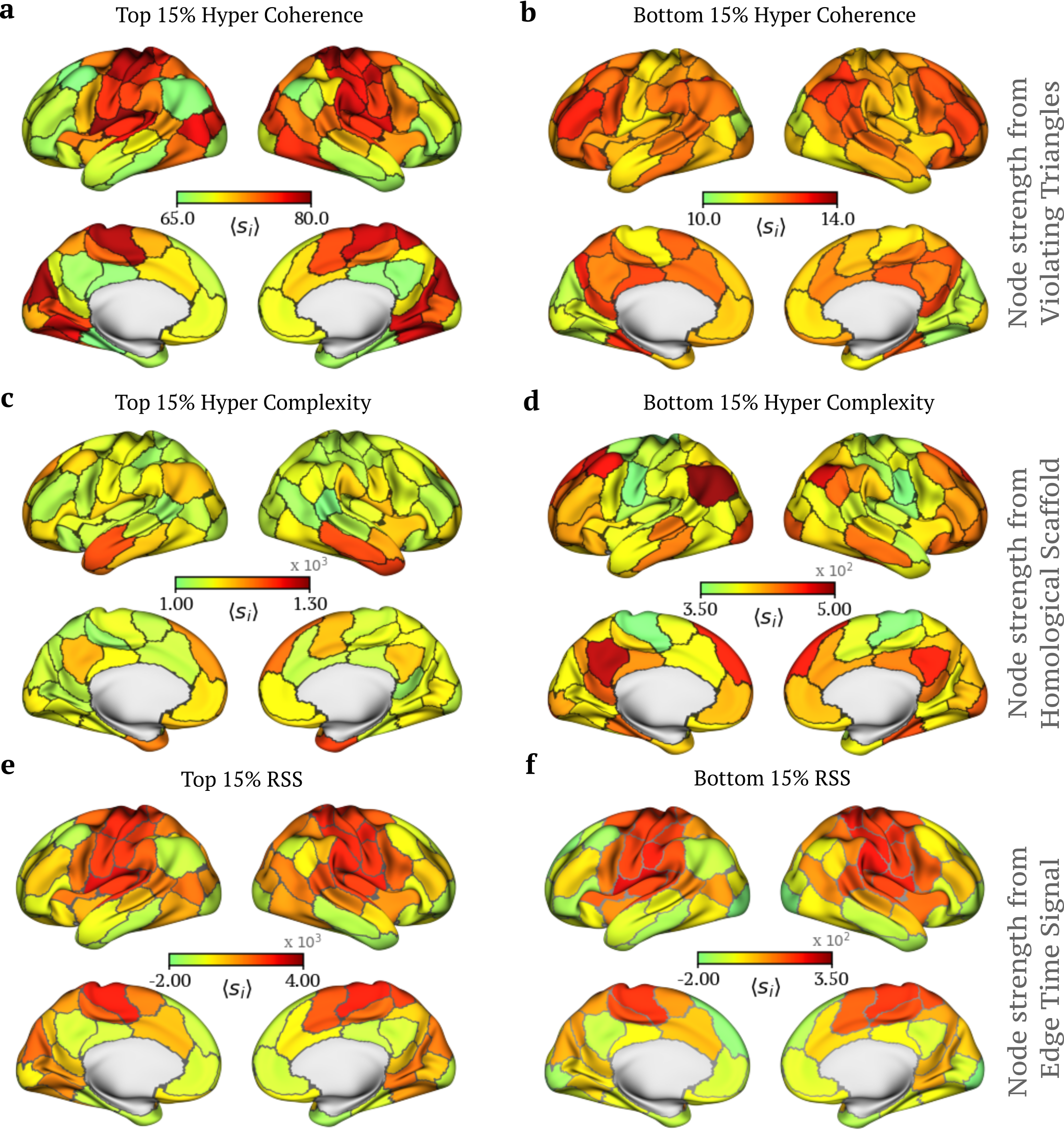}
\caption{\newtext{\textbf{Impact of different group-order representations in brain patterns}. Node strength extracted from the violating triangles $\Delta v$, homological scaffold, and edge time series can be used to track the importance of group-order structures. \textbf{(a-b)} We report the brain map of the nodes involved in higher-order co-fluctuations obtained when isolating the 15\%  high- and low- coherent frames, respectively. \textbf{(c-d)}  We report the brain map of the nodes involved in homological scaffold obtained when isolating the 15\%  high- and low- complexity frames. To compare different group-order representations, we also performed a similar analysis as panels (a-d), yet considering the nodal strength extracted from the edge-wise representation. In particular, \textbf{(e-f)} we map onto the cortical surface the nodal strength extracted directly from the 1-order co-fluctations (i.e. the edge time series),  computed from the 15\%  high- and low-amplitude frames, which have been selected using the RSS statistics. Despite losing part of the information by projecting each group-order representation on a nodal level, we observe that the brain patterns uncovered by top- and low- peaks of the higher-order approach provide different brain configurations compared to edge-wise representation, with the only exception being the top hyper coherence frames (see Fig.~\ref{fig:SI_fig12_correlation_brain_financial_patterns} for all the correlations of activity patterns). Results are averaged over all 100 HCP subjects and scans.}}
\label{fig:SI_fig10_brain_patterns}
\end{figure}

\newtext{Figure~\ref{fig:SI_fig10_brain_patterns} summarizes the results of our comparisons for the brain data. Here, we mapped the node strength extracted from the violating triangles $\Delta v$, the homological scaffold, and the edges time series onto the cortical surface and found various patterns of co-fluctuations. In particular, we first observe that brain patterns uncovered by the 15\% top hyper coherent frames provide quite similar brain configurations as the ones constructed from both top- and bottom-RSS edge-wise representations. By contrast, all the other representations based on higher-order measures provide different brain activity patterns, with the nodal strength of the homological scaffold from the 15\% bottom hyper complexity frames mostly encompassing the Default Mode Network (DMN), see Fig.~\ref{fig:SI_fig10_brain_patterns}(d).}

\begin{table}[hbt!]
\centering
    \begin{tabular}{|c|p{13 cm}|}
        \hline
        \newtext{\textbf{Method}} & \newtext{\textbf{Description}}\\
        \hline
        \newtext{\makecell{$\uparrow \Delta_v$\\($\downarrow \Delta_v$)}} & \newtext{The top (bottom) 15\% of hyper-coherent frames are selected using the Hyper Coherence indicator. For only those frames, we compute the average spatial distributions by projecting the list of violating triangles $\Delta_v$ on a nodal level.}\\
        \hline
        \newtext{\makecell{$\uparrow HC$ \\ ($\downarrow HC$)}} & \newtext{The top (bottom) 15\% of the hyper-complexity frames are selected using the Hyper Complexity indicator. For only those frames, we compute the average nodal strength from the homological scaffold, which is obtained from the persistent homology generators of $H_1$~\cite{petri2014homological_SI}.}\\
        \hline
        \newtext{\makecell{$\uparrow eFC$\\($\downarrow eFC$)}} & \newtext{The top (bottom) 15\% of the  amplitude frames are selected using the RSS measure. From those frames, we then compute the edge functional connectivity matrix~\cite{faskowitz2020edgecentric_SI} and report the nodal strength obtained from such a matrix.}\\
        \hline
        \newtext{\makecell{$\uparrow eTS$\\($\downarrow eTS$)}} & \newtext{The top (bottom) 15\% of the  amplitude frames are selected using the RSS measure. For those frames, we then compute the average activity of the 1-order co-fluctuation (i.e. the edge time series) and report the nodal strength.}\\
        \hline
        \newtext{\makecell{$\uparrow nodal$\\($\downarrow nodal$)}} & \newtext{The top (bottom) 15\% of the frames are selected using a RSS-like statistic but computed on the raw signal. For those frames, we then compute the average activity pattern to obtain nodal maps.}\\
        \hline
    \end{tabular}
\caption{\newtext{\textbf{Summary of the different approaches.} We provide a short description of the methods considered in this study to construct spatial maps from multivariate time series. We remind also that $\Delta_v$ and $HC$ are higher-order measures, $eFC$ and $eTS$ are edge-wise statistics, while $nodal$ corresponds to spatial measures based on the original nodal signal.   }}
\label{table:SI_table_methods_description}
\end{table}

\begin{figure}[t!]
\centering
\includegraphics[width=0.95\textwidth]{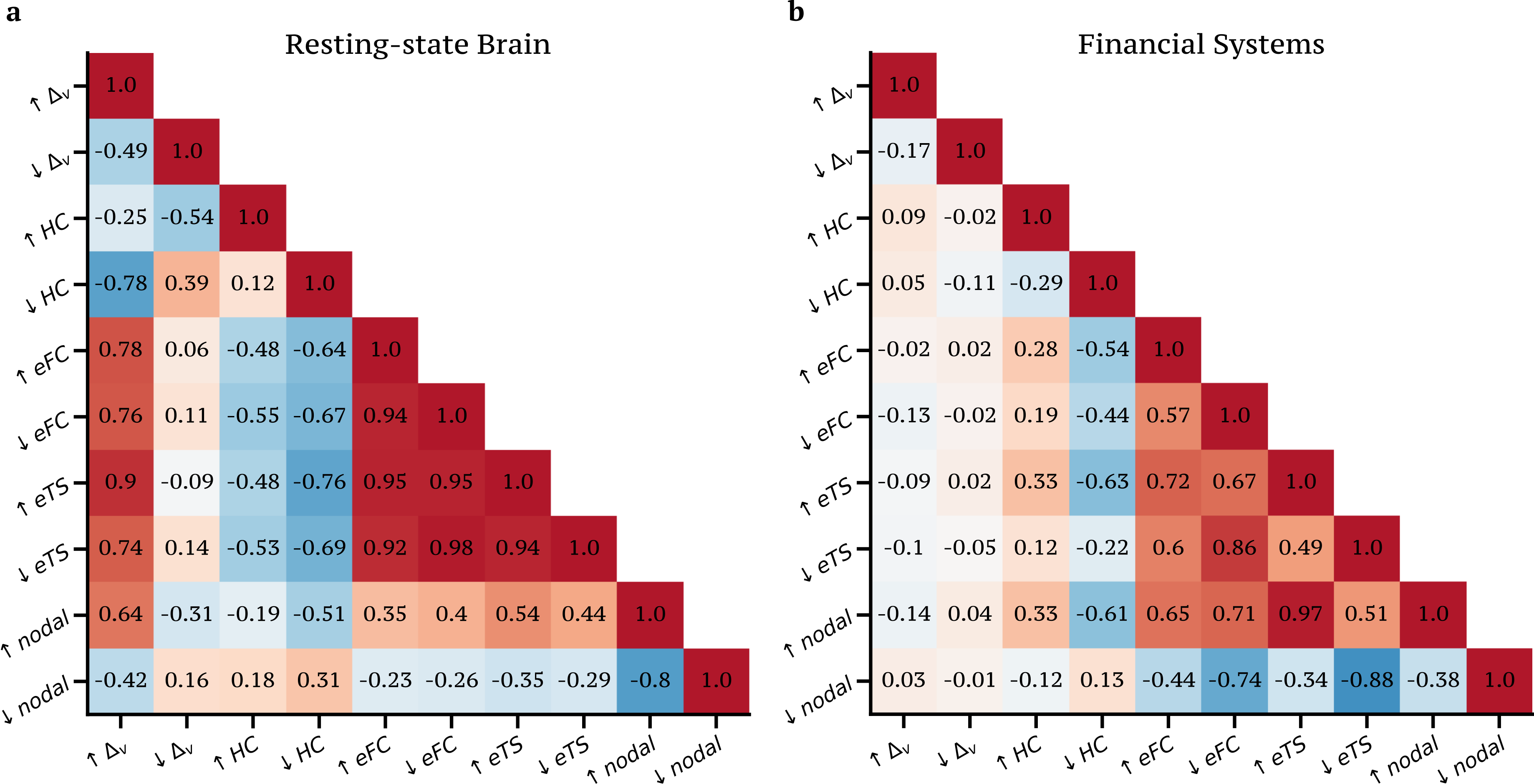}
\caption{\newtext{\textbf{Correlations between group-order representations of brain and financial patterns}. We report the Pearson's correlation coefficient $\rho$ obtained when comparing the nodal projections for \textbf{(a)} brain and \textbf{(b)} financial systems. Here we have that $\Delta_v$ and $HC$ are higher-order measures, $eFC$ and $eTS$ are edge-wise statistics, while $nodal$ corresponds to measures based on the nodal signal. $\uparrow (\downarrow)$ correspond to the 15\% top- (respectively bottom-) peaks of the selected measures. For the sake of clarity, in Table~\ref{table:SI_table_methods_description} we reported a short description of the methods. Interestingly, while the spatial patterns uncover by the 15\% top-peaks of the higher-order approach using the hyper coherence (i.e. $\uparrow \Delta_v$) provide brain patterns very similar to the ones constructed from both top- and bottom-RSS edge-wise representations, all the other three higher-order approaches provide patterns that are not entirely deducible from an edge-wise approach. However, in financial systems we have an opposite view, so that the patterns obtained from the top 15\% hyper coherent frames (i.e. $\uparrow \Delta_v$) is now very dissimilar from all the ones obtained using edge approaches (i.e. $\rho \approx 0$).}}
\label{fig:SI_fig12_correlation_brain_financial_patterns}
\end{figure}

\newtext{Yet, the analyses presented so far provide only a qualitative comparison between the approaches. We therefore quantitatively investigate the spatial patterns identified on both resting-state brain data and financial systems by the various methods and compare them considering nodal projections. \\
For clarity, we first provide in Table~\ref{table:SI_table_methods_description} a summary description of each method, while we report in Figure~\ref{fig:SI_fig12_correlation_brain_financial_patterns}  the matrices encoding all the possible Pearson's correlations between the various spatial distributions for both the brain data and financial data. In particular, by analysing Figure~\ref{fig:SI_fig12_correlation_brain_financial_patterns}(a), we confirm in a quantitative way that the spatial patterns uncovered by the 15\% top-peaks of the higher-order approach using the hyper coherence (i.e. $\uparrow \Delta_v$) provide brain patterns which are very similar to the ones constructed from both top- and bottom-RSS edge-wise representations. Therefore, at least in brain resting data, such representation might lead to similar results as the one obtained using edges.  By contrast, the other higher-order approaches provide patterns that are not entirely deducible from an edge-wise approach and, as such, they might provide new insights into brain dynamics.  However, when analysing the same correlation matrix for financial systems [Figure~\ref{fig:SI_fig12_correlation_brain_financial_patterns}(b)], we surprisingly find a rather different picture. 
In this case, the spatial pattern of the top 15\% hyper coherent frames (i.e. $\uparrow \Delta_v$) is very dissimilar from all those obtained using edge approaches (i.e. $\rho \approx 0$), while the other higher-order approaches still provide patterns that are not captured by any edge-wise approach.  This confirms our intuition that various real-world systems might be affected, in varied ways, by group-interactions at different instants in time.}

\newtext{Finally, by analysing the historical data of epidemic outbreaks in the US, we show that the temporal evolution of the RSS measure cannot be effectively used to classify different infectious diseases. Indeed, with a Random Forest classifier we obtain an accuracy level of around 67\% using a 10-fold cross-validation setting repeated 50 times. See Section \ref{sec:comparison-us-data} for a comparison with classification based on higher-order information.}

\begin{figure}[h!]
    \centering
    \includegraphics[width=0.9\textwidth]{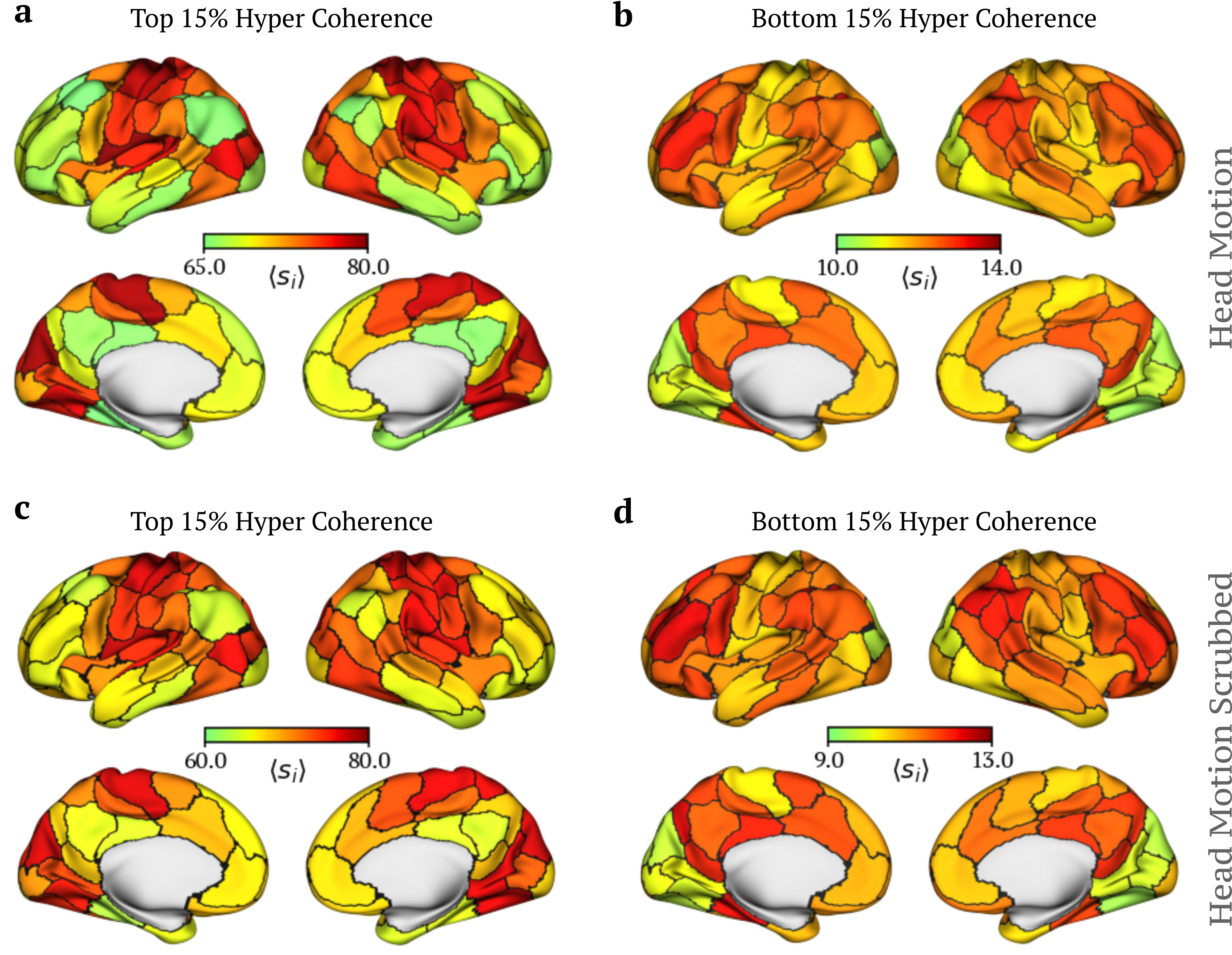}
    \caption{\newtext{\textbf{Effect of head motion in brain maps}. We report the nodal spatial distributions obtained from the violating triangles $\Delta_v$ using the top- and bottom 15\% hyper coherent frames when analysing the fMRI resting-state data in  \textbf{(a-b)} presence and  \textbf{(c-d)} absence of fMRI volumes containing noisy fluctuations (head motion).  When comparing the resulting brain maps we obtain a Pearson's correlation coefficient of $\rho \approx 0.999$.}}
    \label{fig:SI_fig13_noisy_brainmaps}
\end{figure}

\cleardoublepage
\newpage
\newtext{\section{Effect of noisy fluctuations in brain data}}
\newtext{As mentioned in the main text, one of the limitations of our higher-order approach working on single-frame timescales is that it might be affected by noisy fluctuations in the time series, as many of other existing methods~\cite{tagliazucchi2012criticality_SI,liu2013timevarying_SI,esfahlani2020highamplitude_SI}. However, in practice, this problem can be smoothed out by analyzing the statistics of multiple time frames. In this section we confirm that, even in presence of noisy fluctuations in brain data (i.e., head movement), the resulting spatial distribution of the hyper coherence appear to be robust and stable.  More precisely, similarly to Ref.~\cite{jenkinson2002improved_SI},  we considered frames (i.e. fMRI volumes) affected by head motion if the relative Root Mean Square (RMS) between frames was greater than 0.2, i.e. $RMS > 0.2$.\\
    Figure~\ref{fig:SI_fig13_noisy_brainmaps} reports the nodal spatial distributions obtained from the violating triangles $\Delta_v$ using the top- and bottom 15\% hyper coherent frames when analysing the fMRI resting-state data in presence and absence of fMRI volumes containing high head motion. Notably, we obtain a Pearson's correlation coefficient $\rho \approx 0.999$ when comparing the resulting brain maps in presence and absence of fMRI volumes affected by head motion. } 

\cleardoublepage
\newpage

\section{Local higher-order indicators in real-world systems}
\subsection{Role of thresholds in brain maps}
\begin{figure*}[bh!]
    \centering
    \includegraphics[width=1\textwidth]{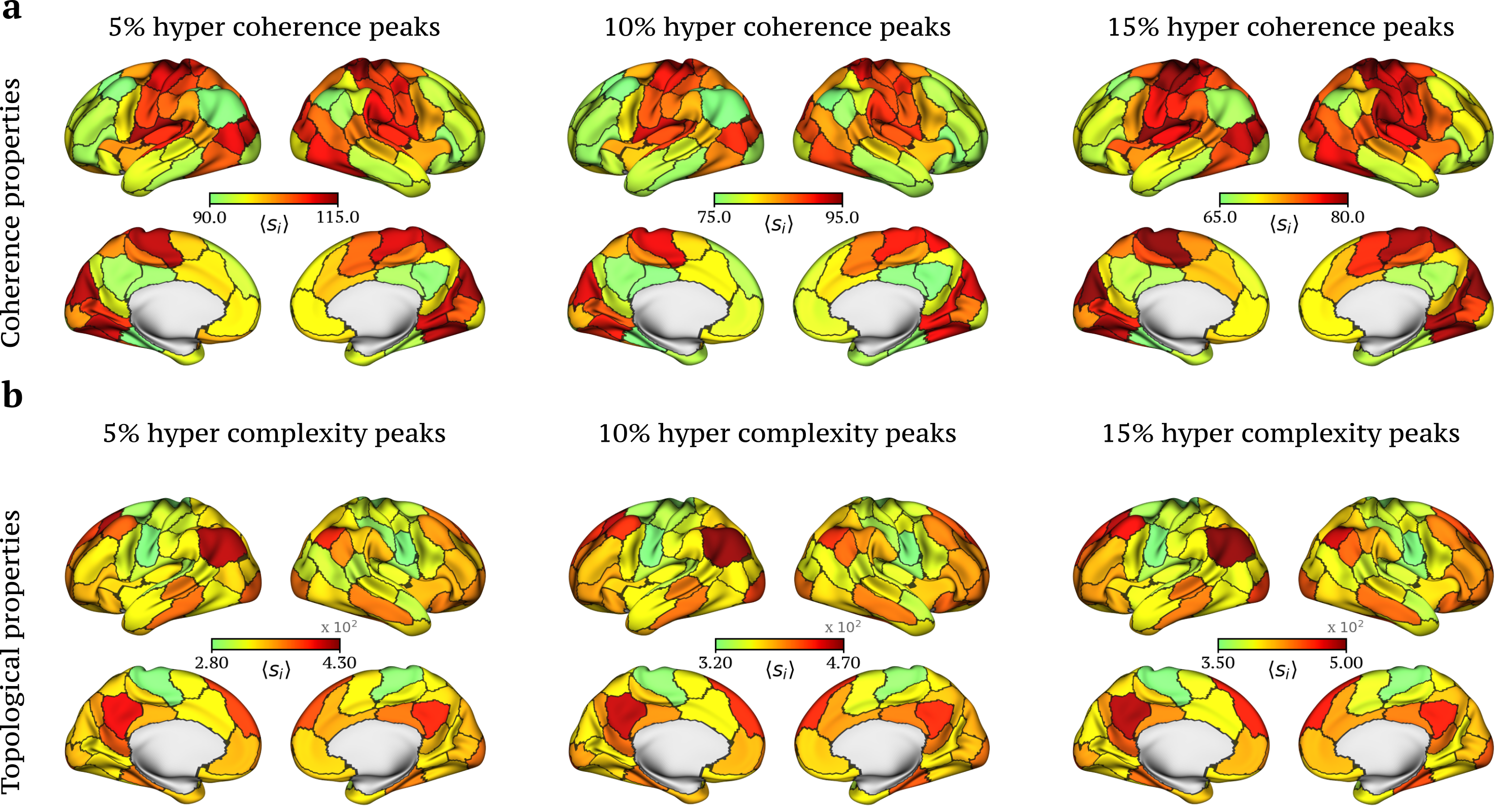}
    \caption{\textbf{Brain maps with different percentage of peaks}. The nodal strength extracted from either the violating triangles $\Delta_v$ or the homological scaffold can be used to track the importance of higher-order structures in time. (\textbf{a}) Brain maps of the nodes involved in higher-order co-fluctuations obtained when respectively isolating the $5,10,15\%$ high-coherent frames, which are those associated with a more synchronized dynamical phase. Interestingly, brain maps are highly consistent across different values of thresholds.  From a topological perspective, the nodal strength extracted from the homological scaffold provides information about 1D loops in the space of co-fluctuations. (\textbf{b}) Brain maps obtained when respectively selecting the 5, 10, 15\% low-hyper complexity frames reveal the consistent activation of the Default Mode Network. Indeed, also in this case,  the brain maps are mostly the same for different threshold values. }
    \label{fig:SI_fig4_brain_maps}
\end{figure*}

In the main text, we have shown that the nodal strength obtained when projecting
either the magnitudes of the violating triangles  $\Delta_v$  or the homological scaffold on a nodal
level might provide interesting information about the group co-fluctuations of certain brain regions. In particular, since our goal was to characterize the
higher-order states with the largest level of synchronization, we isolated only the 15\% high-coherent (resp. low-complex) frames. Yet, in principle, a different percentage of selected peaks might translate into a distinct brain activation maps As a consequence, in Fig.~\ref{fig:SI_fig4_brain_maps} we report the same brain maps as shown in the main text, yet we vary the percentage of peaks selected. 

\newtext{\subsection{Higher-order indicators in the brain's functional networks}}
\newtext{In this section, we consider the local higher-order indicators introduced in the main text to analyze the brain at the level of the 7 functional Yeo networks~\cite{yeo2011organization_SI} and subcortical regions. In particular, we project the normalized magnitude of the violating triangles $\Delta_v$ extracted from the $15\%$ top hyper coherent frames and of the homological scaffold extracted from the $15\%$ low hyper complexity frames at the nodal and edge level. This provides us with information about the most prominent functional networks when isolating a certain percentage of frames both at the level of nodes and edges. Figure~\ref{fig:S1_fig14_brain_yeo_nets}(a) reports the mean hyper coherence within- and between- the 7 functional Networks plus subcortical regions, namely, the Visual (VIS), SomatoMotor (SM), Dorsal-Attention (DA), Ventral-Attention (VA), Limbic (L), FrontoParietal (FP), Default Mode Network (DMN), and subcortical regions (SUBC). While the top boxplots (also shown in the main text) report the mean hyper coherence within the 7 functional Networks, the entries of the matrix encode the mean hyper coherence within- and between- the 7 functional networks and subcortical regions obtained when projecting $\Delta_v$ at the level of their interacting edges. Notably, hyper coherent interactions (i.e. the triangles) are mainly overrepresented in sensorimotor areas. By contrast, when analysing the mean persistence of 1D holes considering the homological scaffold, we find that such loops are highly concentrated around the DMN and FP areas, as shown in Fig.~\ref{fig:S1_fig14_brain_yeo_nets}(b), mirroring the importance of such areas in integrating high and low-order systems.
    \begin{figure}[h!]
        \centering
        \includegraphics[width=0.9\textwidth]{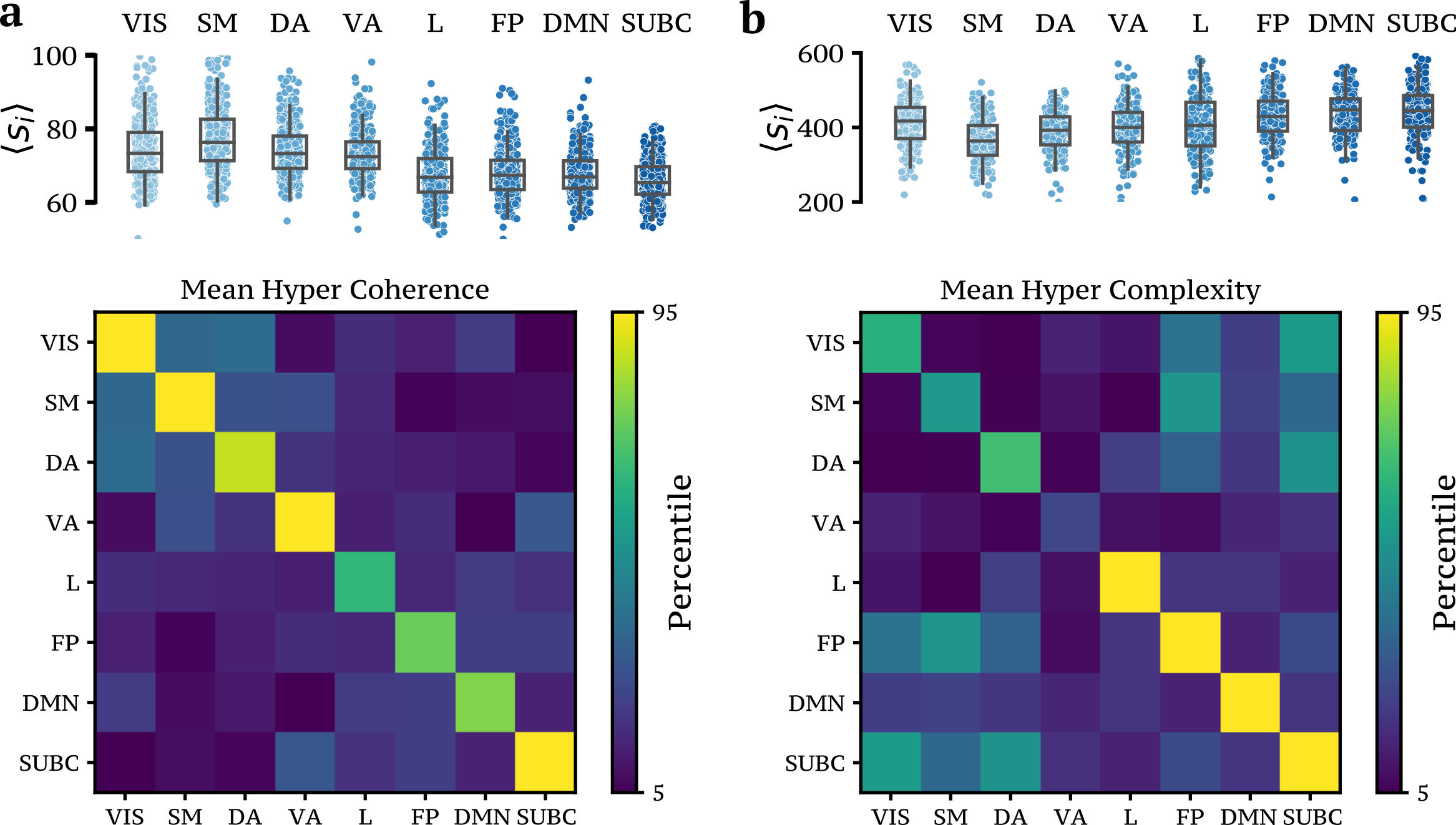}
        \caption{\newtext{\textbf{Mean within- and between higher-order indicators in the seven functional networks.} For the two local higher-order indicators considered in the main text, namely,  \textbf{(a)} the list of violating triangles $\Delta_v$ and  \textbf{(b)} the homological scaffold, we report the mean nodal strength and average edge strength within- and between- the 7 functional Networks and subcortical regions, namely, the Visual (VIS), SomatoMotor (SM), Dorsal-Attention (DA), Ventral-Attention (VA), Limbic (L), FrontoParietal (FP), Default Mode Network (DMN), and subcortical regions (SUBC). When analysing the list of violating triangles $\Delta_v$, we find that activity patterns with emphasized synchronized co-fluctuations mainly reflect sensorimotor areas, whereas the persistence of 1D holes obtained from homological scaffold are situated around the DMN and FP areas.}}
        \label{fig:S1_fig14_brain_yeo_nets}
\end{figure}}

\subsection{Comparison of classifiers for the US historical data}\label{sec:comparison-us-data}
In the main text, we reported the average accuracy score when considering the Support Vector Machine (SVM) method applied to the classification problem of the US historical data of several infectious diseases at the US state-level.

In particular, we have considered the following set of features for the classification task: hyper coherence, the three different contributes of hyper complexity, and the average edge violation. In Table~\ref{table:SI_accuracy_f1_classifier}, we report the scores of several classifiers obtained when considering a 10-fold cross-validation repeated 50 times with different training-test data partitions. The classifiers considered are: Gaussian naive Bayes (Gaussian NB), SVM using a Gaussian radial basis function as kernel (RBF SVM), Decision Tree,  random decision forest (Random Forest), and  k-nearest neighbors algorithm (k-NN, with k=5). Remarkably, the RBF SVM method, together with Random Forest, leads the pack with the highest accuracy of 0.85. All the analysis reported were obtained using the scikit-learn 1.0.2 python library.
\begin{table}[htb!]
    \centering
    \begin{tabular}{|c|cc|}
        \hline
        Classifier & Avg. accuracy & F1 weighted score \\
        \hline
        Gaussian NB & 0.47 & 0.43 \\
        RBF SVM & \textbf{0.85} & 0.85 \\
        Decision Tree & 0.81 & 0.81 \\
        Random Forest & \textbf{0.85} & 0.85 \\
        k-NN & 0.83 & 0.83 \\
        \hline
    \end{tabular}
    
    \caption{\textbf{Comparison of classifier scores}. We report the average accuracy and F1 weighted scores for the classification of the US historical data of several infectious diseases at the US state-level. We consider hyper coherence, avg. edge violations, and the three different contributes of hyper complexity as features for our classification task. The classifiers considered are: Gaussian naive Bayes (Gaussian NB), SVM using a Gaussian radial basis function as kernel (RBF SVM), Decision Tree,  random decision forest (Random Forest), and  k-nearest neighbors algorithm (k-NN, with k=5). }
    \label{table:SI_accuracy_f1_classifier}
\end{table}

\FloatBarrier
\let\oldaddcontentsline\addcontentsline
\renewcommand{\addcontentsline}[3]{}

\let\addcontentsline\oldaddcontentsline

\end{document}